\documentclass{pasj01}
\onecolumn
\usepackage{setspace}
\usepackage{longtable}
\usepackage{subfigure}
\usepackage{epstopdf}
\usepackage{parskip}
\usepackage{booktabs}
\usepackage{multirow}
\usepackage{multicol}
\usepackage{array}
\usepackage{xcolor}
\usepackage[normalem]{ulem}
\usepackage[left,switch,mathlines]{lineno}
\Received{$\langle$12.06.2025$\rangle$}
\Accepted{$\langle$30.10.2025$\rangle$}
\Published{$\langle$publication date$\rangle$}

\begin{document}
\title{Constraining the Corona Geometry of Cyg X-1 with Broad Band Spectrum and Polarimetric 
Analysis Based on Observations in May 2022}

\author{Sixuan \textsc{zhang}\altaffilmark{1,4}}%
\altaffiltext{1}{Hiroshima Astrophysical Science Center, Hiroshima University, 1-3-1 Kagamiyama,
Higashi-Hiroshima, Hiroshima 739-8526 Japan}
\email{d211678@hiroshima-u.ac.jp}

\author{Tsunefumi \textsc{mizuno}\altaffilmark{1}}
\email{mizuno@hiroshima-u.ac.jp}

\author{Tomohisa \textsc{kawashima}\altaffilmark{2}}
\altaffiltext{2}{Department of Engineering for Future Innovation, National Institute of Technology,
Ichinoseki College, Ichinoseki, Iwate 021-8511, Japan}

\author{Chris \textsc{done}\altaffilmark{3}}
\altaffiltext{3}{Department of Physics, University of Durham, Durham DH1 3LE, UK}

\author{Yasushi \textsc{fukazawa}\altaffilmark{4}}
\altaffiltext{4}{Graduate School of Advanced Science and Engineering, Hiroshima University, 1-3-1
Kagamiyama, Higashi-Hiroshima, Hiroshima 739-8526, Japan}

\author{Hiromitsu \textsc{takahashi}\altaffilmark{4}}

\author{Ryusei \textsc{komine}\altaffilmark{5}}
\altaffiltext{5}{Center for Computational Sciences, University of Tsukuba, 1-1-1 Tennodai, Tsukuba,
Ibaraki 305-8577, Japan}

\author{Koudai \textsc{takebayashi}\altaffilmark{5}}

\author{Ken \textsc{ohsuga}\altaffilmark{5}}


\maketitle

\begin{abstract}
Cygnus X-1 (Cyg X-1) exhibited a low hard state in 2022, observed by several missions. The
\textit{IXPE} reported that the polarization angle (PA) is aligned with the radio jet and
gave a polarization degree (PD) approximately 4 times higher than the general expectations
of $1\%$ through the analysis of the time-integrated data with a simple spectral model,
indicating that the disk inclination is higher than a canonical value of about $30^{\circ}$.
Many subsequent theoretical studies employed a non-standard model to explain this high PD.
Here, we revisit the disk/corona spectrum through a detailed joint analysis using \textit{IXPE},
\textit{NuSTAR}, and \textit{NICER} data. By investigating the time variability of the
spectrum, we find that the two-Comptonization components model can better reproduce the
data than the one-Comptonization component model originally adopted. We observed a lower
disk photon temperature of about 0.15 keV. Detailed simulation suggests that lowering
the disk temperature by a factor of 2 increases the PD by \textcolor{black}{roughly 2 percentage
points in the \textit{IXPE} 2--8 keV band for a slab-like corona geometry}, helping to reconcile
the observed high PD with theoretical predictions. \textcolor{black}{However, The simulated PDs are
still significantly lower than the observed ones - even for a rather high $60^{\circ}$ inclination.}
We also investigated the polarization properties of a simple wedge-shaped corona with a truncated
disk and a sandwiching slab corona. We find that the slab corona predicts an apparent energy
dependence in PD while PA remains constant in the \textit{IXPE} band, in agreement with the
observed polarization. Therefore, we suggest that Cyg X-1 in 2022 May exhibits a
two-Comptonization coronal emission with different optical depths, and the hard one is in a
sandwiching slab geometry. We also discuss how the polarization is affected by other
parameters of the black hole (BH) and the corona.
\end{abstract}

\section{Introduction}

\hspace{2.5em} Accretion disks around black holes (BH) are often classified as different states,
low hard state (LHS), high soft state (HSS), and very high state (VHS), or steep power law
state for some BH systems, corresponding to different accretion rates (\cite{Done2007}). 
In HSS, the accretion disk is closed to the BH resulting in high inner disk temperature, and 
the thermal emission from an optically thick disk will dominates the spectrum under the 
condition of high accretion rate (\cite{Shakura1973}; \cite{Remillard2006}); On the 
contrary, the disk might be truncated with lower disk temperature in LHS and the spectrum will 
be dominated by Compton scattering, which can be approximately described by a simple power-law 
(PL) model with a high-energy cutoff due to the low accretion rate; More over, some sources (i.e.
Cyg X-1) may also exhibit a so-called VHS in which both thermal disk and corona emissions are
strong, corresponding to a very high accretion rate (\cite{Remillard2006}). 

Cyg X-1 is one of the brightest sources in the sky, which has been observed and studied for
a long time in multiple wave bands. It is a binary system consisting of the first widely
accepted BH of $21.2 \pm 2.2 M_\odot$ and an O-type supergiant around 40$M_\odot$
(\cite{Walborn1973}). Studies also suggest that the BH is spinning rapidly with the dimensionless
spin factor $a > 0.92$ (\cite{Gou2014}). This high-mass X-ray binary system (HMXB) is
$2.22^{+0.18}_{-0.17}$ kpc away and has an orbital period of around 5.6 days (\cite{Miller2021}).
It is also known as a microquasar, since a radio jet is informed by radio observations
(\cite{Stirling2001}). X-ray observations will help to reveal the geometry of the accretion
disk and corona.

Hybrid models (disk radiation, Comptonization, and disk reflection) were applied to analyze
the Cyg X-1 spectrum in both the soft and hard states. The soft state of Cyg X-1 is often
attributed to a hot, optically thick accretion disk. The disk extends down to the innermost
stable circular orbit (ISCO) and produces thermal radiation, and the hot corona produces the
hard tail. The spectrum is dominated by thermal disk radiation and Comptonization by hybrid
thermal/nonthermal plasma (\cite{Gierlinski1999}). In contrast, the hard state of Cyg X-1 can
be described by a model consisting of a truncated disk that produces cool thermal disk
radiation and a hot plasma that produces Comptonized emission. However, the shape of the
Compton corona is still in debate. In the cone-shaped corona model, the hot plasma is assumed
to extend along the spin axis of the BH, while the corona is spherical and located above and
below the BH in the so-called lamp-post model (\cite{Beheshtipour2022}). They also applied
a wedge-shaped and a sandwiching slab corona to Cyg X-1, where the corona extends along the
disk plane rather than the spin axis. Moreover, some studies also argue that the disk can
reach the ISCO even in a low/hard state (\cite{Parker2015}), which is challenged by following
study (\cite{Basak2017}). New studies also give evidence for two different Comptonization
components (\cite{Basak2017}; \cite{Kawamura2022}). For example,
\citet{Makishima2008}
suggested that Cyg X-1 exhibits a two-Comptonization with different optical depths
in the hard state in 2005. The different disk and corona geometries show minor differences
in the spectrum, making it difficult to distinguish them by spectral analysis. With the help
of polarimetric observation in X-rays, we may finally constrain the disk and corona geometry.

The Imaging X-ray Polarimetry Explorer (IXPE) mission (\cite{Weisskopf2022}), launched on
9 December 2021, is the first mission dedicated to detecting polarizations since OSO-8
(\cite{Weisskopf1977}), and gives us the opportunity to measure the polarization at high
precision in the 2-8 keV band. It is much more sensitive than OSO-8 and allows us to detect
the source with the same sensitivity with an observation time 100 times shorter. Cyg X-1 was
observed by IXPE twice in 2022, between May 15 and 21 and between June 18 and 20
(\cite{Krawczynski2022}; hereafter referred to as the discovery paper). The first observation
has a total net exposure time of around 242 ks, while the second observation is around 86 ks,
about one-third of the first observation. The two observations show that Cyg X-1 was in a hard
state.   

The expected polarization degree (PD) of Cyg X-1 in the 2-8 keV \textit{IXPE} band is around
$1\%$ , depending on the state (\cite{Schnittman2009}; \cite{Schnittman2010};
\cite{Beheshtipour2022}) with the inclination angle $\iota =27^{\circ}.5\pm 0^{\circ}.8$
inferred from optical observations (\cite{Miller2021}). With the combination of 3 detectors,
\textit{IXPE} detected PD of $4.0 \pm 0.2\%$ on average in the 2 -- 8 keV band and also
suggested a marginal energy dependency of PD. The polarization angle (PA), the same as the
electric vector position angle, is $-20^{\circ}.7\pm 1^{\circ}.4$, where the positive value
represents a position angle east of north. The PD and PA of the second observation are
$3.76 \pm 0.24\%$ and $-26^{\circ}.2 \pm 1^{\circ}.8$, respectively. The PA agrees well
with the radio jet that is expected to be perpendicular to the disk plane. This helps us
to rule out the lamp-post model in agreement with previous results of \textit{PoGO+} in the
hard X-ray (\cite{Chanuvin2018}). Instead, the PA suggests a truncated disk with a disk
Comptonization geometry or a sandwiching slab corona geometry, where the corona is extended
along the disk plane. However, the high PD cannot be explained by these models with the
inclination angle obtained from optical observations.

Various theories assuming non-standard effect have been proposed to explain the detected
high PD of Cyg X-1, such as higher inclination, an out-flowing corona, a hollow jet, etc.
With an inclination angle larger than $45^{\circ}$, we may observe a $4\%$ PD with the models
assumed in the discovery paper (disk radiation, one-Comptonization component, and reflection
components). The Lense–Thirring Precession of Quasi-periodic oscillations (QPOs) of Cyg X-1
may be another explanation, where PD is modulated by the QPO frequency with the changing
cycle of the viewing angle (\cite{Ingram2015}). Alternatively, a hot Compton corona with
outflow speeding at $50\%$ of the speed of light (\cite{Poutanen2023}) or a hollow-cone
geometry jet can also explain the high PD (\cite{Dexter2023}). We note that the \textit{IXPE}
Cyg X-1 discovery paper analyzed the time-integrated data with a simple one-Comptonization
component model. The true spectrum will likely be more complex and affect the expected
polarization. This paper seeks to construct an improved model to interpret the observed
spectrum and reassess the corona geometry. In the following sections, we will explain
the data and preparatory analysis to establish the spectral model in Section 2. In Section 3,
we describe the spectral-polarimetric analysis and present the results. Finally, we will
discuss polarization properties based on a dedicated simulation in Section 4 and provide the
conclusions in Section 5.

\section{Model Development}

\subsection{Data Sets}
We analyzed the data of Cyg X-1 observed by \textit{IXPE}, \textit{NuSTAR}, and
\textit{NICER} in 2022.

The \textit{IXPE} data is based on the level 2 data between 2022 May 15 and 21 (hereafter
observation 1) with a total exposure time of 242 ks, which is now available in NASA's
HEARSARC data archive\footnote{\texttt{https://heasarc.gsfc.nasa.gov/docs/archive.html}}
as well as the additional observation one month later between June 18 and 22 (observation 2)
with a total exposure time of 86 ks. The source region is defined as a circular region
with a radius of $150^"$, centered on RA $19^h58^m21.67^s$ and Dec $+35^{\circ}12'05.77''$.
The background region is also a circular region of the same radius, but is $300^"$ away from
the source. The level 3 data of \textit{IXPE}, including region-selected events, spectra, etc.
were generated by \texttt{xpselect} and \texttt{xpbin} using the software developed for
\textit{IXPE} called \texttt{ixpeobssim} (version v31.0.1). We used \texttt{HEAsoft}
(version v6.33.1) for spectrum analysis. \textit{IXPE} is sensitive between 2 and 8 keV
with an energy resolution of $\sim 40\%$ at 2 keV, which is insufficient for spectrum
modeling. Therefore, we also used data from \textit{NICER} and \textit{NuSTAR} for
spectral analysis.

The Nuclear Spectroscopic Telescope Array (\textit{NuSTAR}) (\cite{NuSTAR2013}), which
is sensitive between 3 and 78 keV, observed Cyg X-1 three times between May 15 and 18,
2022 with a total exposure time of 42 ks. \textit{NuSTAR} has high sensitivity above 20 keV,
which is crucial to constrain the Comptonization parameters, especially the electron
temperature $kT_e$ of the corona. We used the \texttt{NuSTARDAS} software (version v1.97)
of the \texttt{HEAsoft} package.

The Neutron star Interior Composition Explorer (\textit{NICER}), sensitive between 0.2
and 12 keV, observed Cyg X-1 between May 15 and 21, with a total exposure time of 87 ks.
It covers almost the entire \textit{IXPE} observation time and has a better energy
resolution (85 eV at 1 keV and 140 eV at 6 keV), which is crucial to constrain the disk
emissions that produce the seed photon (\cite{Gendreau2016}). We used the \texttt{NICERDAS}
software (version v9.0) of the \texttt{HEAsoft} package to process the \textit{NICER} data.

\subsection{Light Curves}
\begin{figure*}[h]
    \centering
    \subfigure{\includegraphics[width=0.45\textwidth]{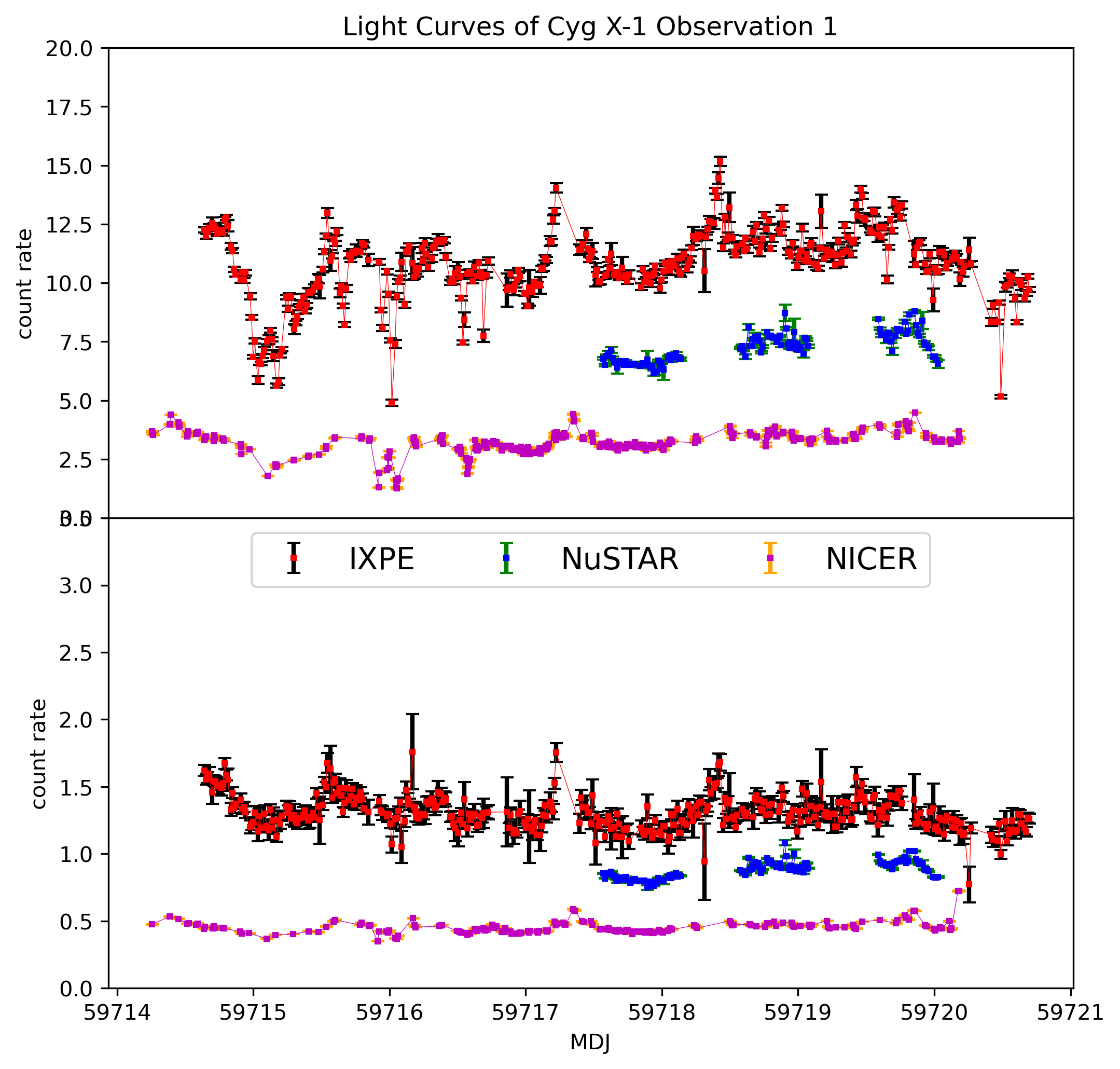}}
    \subfigure{\includegraphics[width=0.45\textwidth]{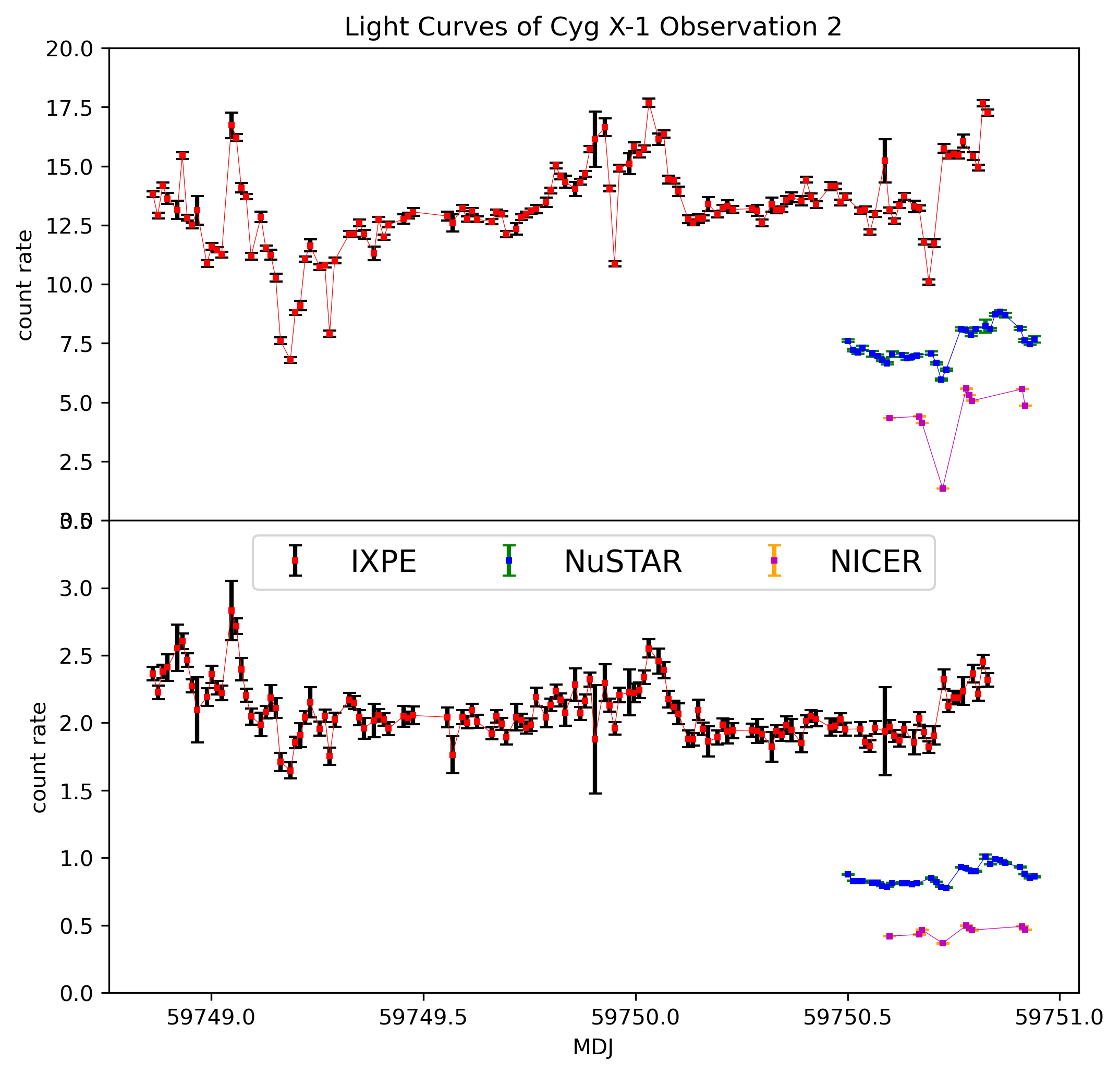}}
    \caption{\textbf{Light curves of observation 1 and 2.} We exhibit the light curves of Cyg X-1 
    observed by, \textit{IXPE} (red), \textit{NuSTAR} (blue) and \textit{NICER} (pink). The left
    panel shows the light curves of observation 1 and the right panel shows those of observation
    2. The upper and lower panels are the light curves below and above 4 keV. Scale factors are
    applied to make count rates of different missions compatible with each other and that of 
    \textit{IXPE} is fixed to 1.
    {Alt text: Four light curves for Cyg X-1 observations in 2022.}}
    \label{LC compare}
\end{figure*}

The average count rate of observation 2 is higher than that of observation 1 by approximately $20\%$
\textcolor{black}{below 4 keV}, \textcolor{black}{whereas it is about a factor of 2 higher above 4 keV
(see figure~\ref{LC compare}), suggesting that} observation 2 is slightly harder. 

Due to the limited co-observation time of
\textit{IXPE} with \textit{NICER} and \textit{NuSTAR}, we exclude observation 2 from further
analysis. The two dips at the beginning and end of observation 1 might correspond to the same
orbital phase, since the time difference is close to the orbital period of 5.6 days. 

The light curves below 4 keV exhibit significant variability, whereas those above 4 keV remain
relatively stable, particularly during observation 1. This behavior is most likely attributable
to changes in absorption; however, intrinsic spectral variations may also contribute. A suitable
spectral model should account for the observed changes using physically reasonable parameters.
In the following sections, we will explore the spectrum in greater detail.

\subsection{Comparison between Single and Double Comptonization Components Model}
\begin{figure}[h]
    \centering
    \includegraphics[width=0.9\linewidth]{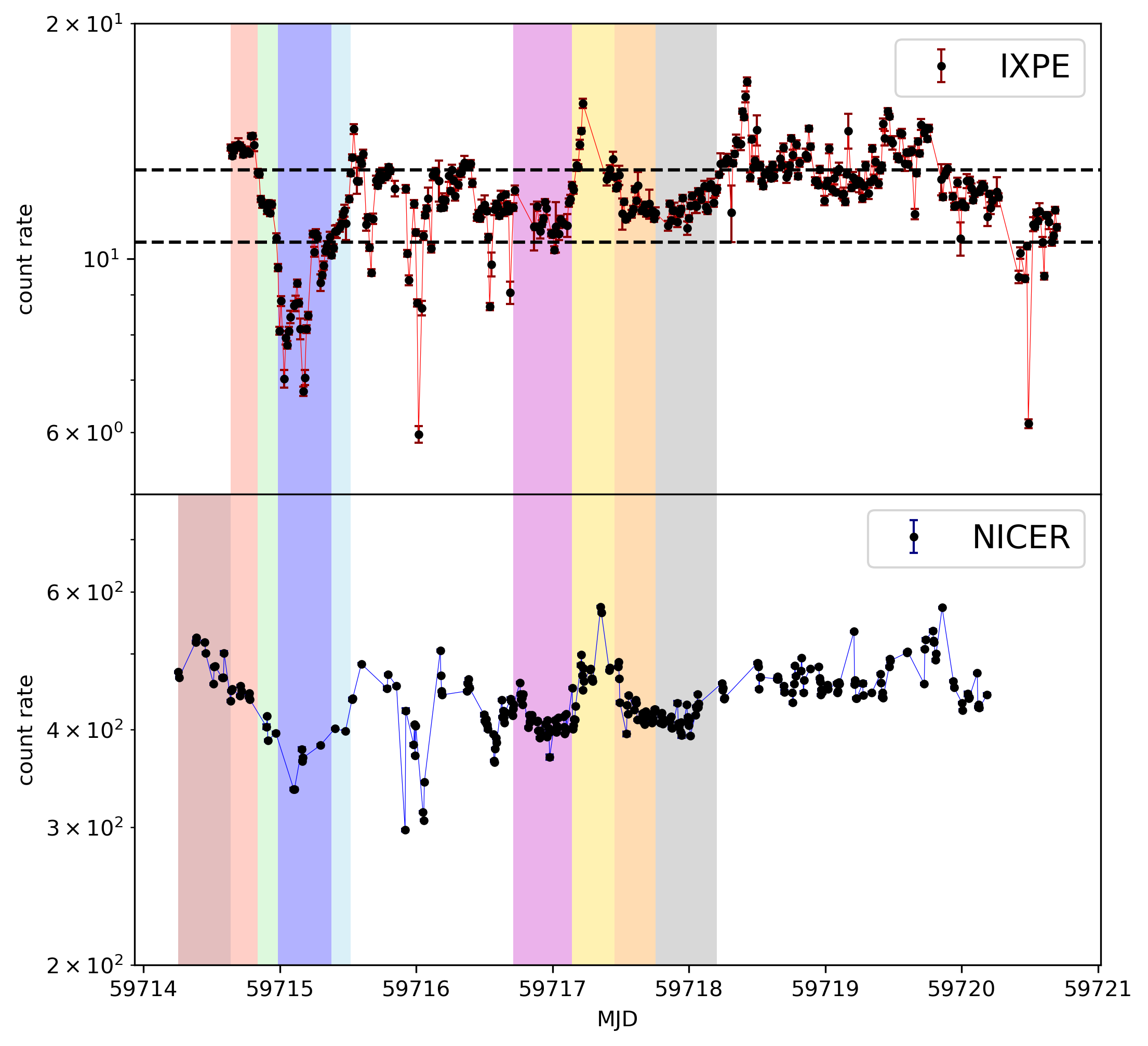}
    \caption{\textbf{Light curves of Cyg X-1 observation 1 of \textit{IXPE} and \textit{NICER}.}
    The upper panel shows the light curve of \textit{IXPE} in 2 -- 8 keV and the lower panel shows 
    that of \textit{NICER} in 0.3 - 15 keV band \textcolor{black}{The shades with different colors
    are the selected time periods closed to the negative dip and positive spike in the \textit{IXPE}
    light curve (summarized in table~\ref{name}).}
    {Alt text: Two light curves of Cyg X-1 observation 1.}}
    \label{LCnicerixpe}
\end{figure}

We compared the detailed light curves of \textit{IXPE} and \textit{NICER} in
figure~\ref{LCnicerixpe}. The significant variation can be due either to the change in
absorption or the intrinsic evolution of the emission from the disk and corona. To
investigate the cause of the change in the spectrum, we first defined two thresholds for the
\textit{IXPE} light curves, as shown by the black dashed line in the upper panel of
figure~\ref{LCnicerixpe}:
\begin{itemize}
    \item The data above the upper threshold are the high flux data;
    \item The data between the upper and lower thresholds are the medium flux data;
    \item The data below the lower thresholds are the low flux data.
\end{itemize}

Since \textit{NICER} has much better photon statistics than \textit{IXPE}, we analyzed the
\textit{NICER} data that perfectly matched the \textit{IXPE} data in developing the model.
Eight time periods are selected, and we label them from i to viii (table~\ref{name} and
figure~\ref{LCnicerixpe}). Time periods 3 and 6 correspond to the negative dip and the
positive spike, respectively.

\begin{table*}[]
    \centering
    \tbl{Definatin of time periods for the preparatory analysis based on the light curves
    in figure~\ref{LCnicerixpe}. To be aware that the leftmost time period shown in the
    \textit{NICER} light curve (dark red in figure~\ref{LCnicerixpe}) are labeled as 0.}{%
    \begin{tabular}{|c|c|c|c|c|c|c|c|c|}
    \hline \hline
       color  &  red  &  green & blue & cyan & pink & yellow & orange & gray \\
    \hline
       number &   i   &    ii   &   iii  &  iv   &   v  &    vi   &   vii    &  viii   \\
    \hline\hline
    \end{tabular}}
    \label{name}
\end{table*}

We fit the data with the one-Comptonization component model, which is taken from the 
discovery paper, and expressed as,
\begin{equation}
    \centering
    \mathrm{TBABS} * (\mathrm{diskBB} + \mathrm{NthComp} + \mathrm{relxillCp} + \mathrm{xillverCp})
    \label{sigcomp model}
\end{equation}
We also employed the two-Comptonization components model, which is written as,
\begin{equation}
    \centering
    \mathrm{TBABS} * (\mathrm{diskBB} + \mathrm{NthComp} + + \mathrm{NthComp}
    + \mathrm{relxillCp} + \mathrm{xillverCp}).
    \label{twocomp model}
\end{equation}

There, $\texttt{TBABS}$ is the absorption model (\cite{Wilms2000}), \texttt{diskbb} is the disk
radiation model that gives the spectrum of an accretion disk consisting of multiple blackbody
components (\cite{Mitsuda1984}; \cite{Makishima1986}), and $\texttt{NthComp}$ is the
comptonization component that produces a thermally comptonized continuum spectrum
(\cite{Zdziarski1996}). $\texttt{relxillCp}$ and $\texttt{xillverCp}$ are the relativistic
and standard reflection models, respectively (\cite{Dauser2014}; \cite{Garcia2014}). 

We fix the electron temperature of the Comptonization and the reflection components to be
$94$ keV and $140$ keV, respectively, the same as the discovery paper. We also fix the
\texttt{diskBB} component when fitting with the two-Comptonization components model, since
it degenerates with the soft Comptonization component. The inner disk temperature $kT_{in}$
is fixed at 0.118 keV, obtained from period 0 where the photon statistic is the best. The
normalization is fixed to 3790, taken from the discovery paper, to simplify the analysis.
Time period iii suffers from very high absorption and we found that the spectrum cannot be
well explained by both models. We indicate that this is caused by the large changes in
absorption within the time periods and decided to ignore the data.

\begin{figure*}[h]
    \centering
    \subfigure{\includegraphics[width=0.4\textwidth]{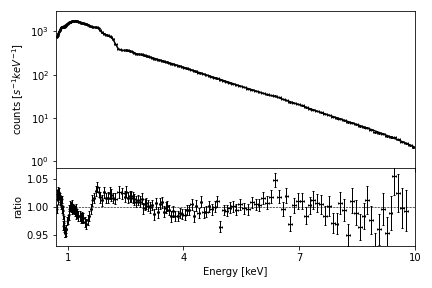}}
    \subfigure{\includegraphics[width=0.4\textwidth]{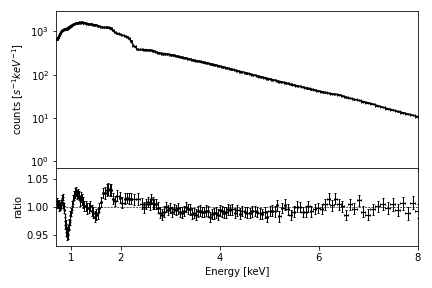}}
    \subfigure{\includegraphics[width=0.4\textwidth]{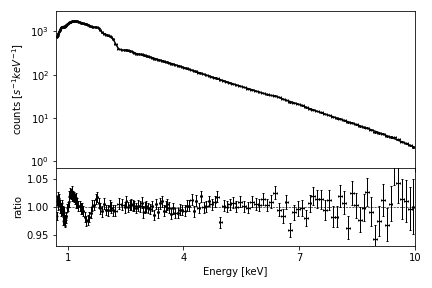}}
    \subfigure{\includegraphics[width=0.4\textwidth]{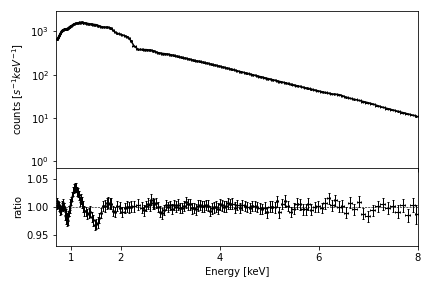}}
    \caption{\textbf{The fitting results of time periods i and vi with different models.} The
    upper panel shows the fitting spectrum and data-to-model ratio of one-Comptonization
    component model for period i (left) and vi (right), respectively. The lower panel presents
    those of two-Comptonization components model. Systematic uncertainties are already taken
    into account\footnote{\texttt{
 https://heasarc.gsfc.nasa.gov/docs/nicer/data\_analysis/nicer\_analysis\_tips.html}}.
 {Alt text: Four spectrum and data-to-model ratio plots of the selected data of
 Cyg X-1 observation 1.}}
    \label{NICERcompare26}
\end{figure*}

We analyzed the other 7 time periods and examined the obtained spectral parameters and the
fit residual. We found that intrinsic changes in the emission mainly cause the spectral
changes. We show the fitting results of time periods i (around the negative dip) and vi
(the positive spike) in table~\ref{parameters} and figure~\ref{NICERcompare26} as an
example. Both the one-Comptonization component model and the two-Comptonization components
model can roughly reproduce the spectrum, but the two-Comptonization components model gives
a smaller $\chi^2$ by approximately 100. In the data-to-model ratio (in
figure~\ref{NICERcompare26}), while both models exhibit similar structures around 1 keV,
the one-Comptonization component model exhibits distinct residuals near 2 and 6 keV.
Other time periods similarly exhibit pronounced residual structures in the data-to-model
ratio at around 2 and 6 keV when modeled with the one-Comptonization component, and
they are effectively mitigated by adopting the two-Comptonization components model. 
Therefore, we conclude that the two-Comptonization components model, which provides a
significantly better fit to the data, is required for a reliable interpretation of the
observed spectra.

\begin{table*}[htb]
    \centering
    \tbl{\textbf{Best fitting parameters comparison of time period i and vi between one
    Comptonization component model and two-Comptonization components model}}{%
    \begin{tabular}{llcccc}
    \hline\hline
    \multirow{2}*{Model} & \multirow{2}*{parameter (unit)} &   
    \multicolumn{2}{c}{one-Comptonization component model}
    & \multicolumn{2}{c}{two-Comptonization components model}\\
      & & i & vi & i & vi \\
    \toprule
    \hline
    \texttt{$TB_{abs}$} & $n_H (10^{22}\ cm^{-2})$ & 
                    $0.474_{-0.009}^{+0.010}$ & $0.484_{-0.010}^{+0.010}$ & 
                    $0.595_{-0.010}^{+0.099}$ & $0.573_{-0.010}^{+0.010}$ \\
    \hline
    \multirow{2}*{\texttt{$diskBB$}} & $T_{in}$ (keV) & 
                    $0.241_{-0.007}^{+0.008}$ & $0.232_{-0.009}^{+0.010}$ & 
                    0.119 (fix) & 0.119 (fix) \\
                                     & N ($10^4$) [$r_g$] & 
                    $6.71_{-1.26}^{+1.59}$ [9.64] & $5.81_{-1.35}^{+1.72}$ [8.95] & 
                    0.379 (fix) & 0.379 (fix) \\
    \hline
    \multirow{2}*{\texttt{$NthComp_s$}} & $\Gamma$ & 
                     - & - & $4.56_{-0.13}^{+0.13}$ & $4.32_{-0.17}^{+0.17}$ \\
                               &    N     &
                     - & - & $2.24_{-0.10}^{+0.10}$ & $1.57_{-0.08}^{+0.08}$ \\
    \hline
    \multirow{2}*{\texttt{$NthComp_h$}} & $\Gamma$ & 
                     $1.70_{-0.01}^{+0.01}$ & $1.66_{-0.01}^{+0.01}$ & 
                     $1.66_{-0.01}^{+0.01}$ & $1.62_{-0.01}^{+0.01}$ \\
                               &    N     &
                     $1.88_{-0.03}^{+0.03}$ & $1.82_{-0.02}^{+0.02}$ & 
                     $1.81_{-0.04}^{+0.04}$ & $1.74_{-0.04}^{+0.04}$ \\
    \hline
    \texttt{$relxillCp$} & N ($(10^{-2})$) & 
                     $4.56_{-0.47}^{+0.46}$ & $6.02_{-0.44}^{+0.41}$ &  
                     $4.82_{-0.47}^{+0.46}$ & $5.79_{-0.44}^{+0.43}$ \\
    \hline
    \multicolumn{2}{c}{$\chi^2$/d.o.f} & 830.03/928 & 645.05/728 & 761.82/922 & 517.04/728 \\
    \hline
    \bottomrule
    \end{tabular}}
    \label{parameters}
\end{table*}

\section{Results of Spectrum and Polarimetric Analysis}

\subsection{Data Selection and Constrain on Corona Temperature }

\begin{figure}[h]
    \centering
    \includegraphics[width=0.95\linewidth]{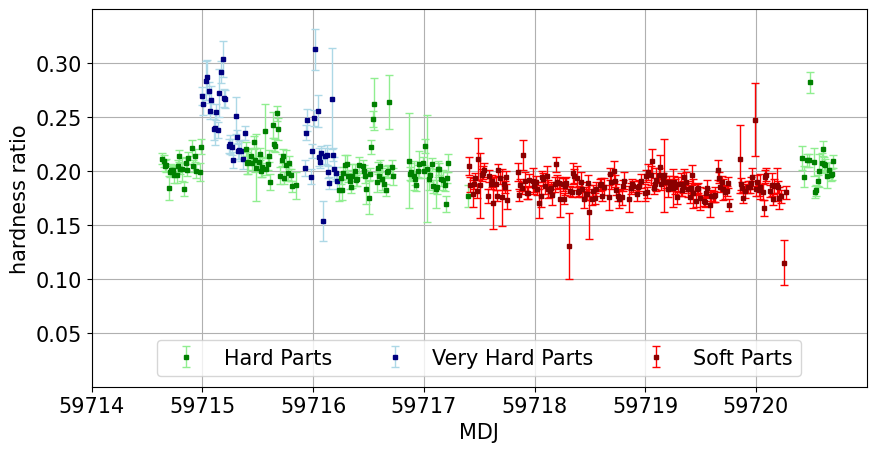}
    \caption{\textbf{The hardness ratio of \textit{IXPE observations 1.}} The figure shows
    the hardness ratio of observation 1. The green and red parts are the relatively hard
    and soft parts, respectively. The blue parts represent the very hard part, corresponding
    to the negative dips shown in figure~\ref{LCnicerixpe}.
    {Alt text: Hardness ratio plot of Cyg X-1 observation 1.}}
    \label{hardness}
\end{figure}

Since Cyg X-1 varied significantly during observation, analyzing the time-integrated spectrum
might mix up different spectral shapes (different contributions from the disk, soft Comptonization,
and the hard Comptonization) and introduce inadequate parameters. Therefore, the hardness ratio
was calculated using the \textit{IXPE} data to separate the data into each time period with a
similar spectrum. The hardness ratio is defined as the count rate below 4 keV over the count
rate above 4 keV. Because absorption affects the low-energy count rate, we ignore the data below
2.5 keV. The obtained hardness ratios are shown in the figure~\ref{hardness}. The blue segments
exhibit a very high hardness ratio. They correspond to the negative dips and are likely to be
affected by strong absorption. Therefore, we will exclude the data from the blue parts. 

\textit{IXPE}, \textit{NICER}, and \textit{NuSTAR} observed Cyg X-1 simultaneously three times
in observation 1 (see figure~\ref{LC compare}), and hereafter we label the first simultaneous
observation as time period A. The relatively hard and soft parts of observation 1 are labeled
as time periods B and C, respectively. We used time period A to simultaneously fit the data of
\textit{IXPE}, \textit{NICER}, and \textit{NuSTAR} to determine the electron temperature of the
hard Comptonization corona. Meanwhile, time periods B and C are used for spectrum and
polarimetric analysis. The data we define are summarized in table~\ref{data}.
\begin{table*}[htbp]
    \centering
    \begin{tabular}{c >{\raggedright\arraybackslash}p{12cm}}
        \toprule \toprule
        Time Periods & \multicolumn{1}{c}{Caption} \\
        \midrule
        \multirow{2}{*}{A} &
        The first simultaneous observation of Cyg X-1 by \textit{IXPE},
        \textit{NuSTAR}, and \textit{NICER} shown in figure~\ref{LC compare}. \\
        \addlinespace[2pt]
        \multirow{2}{*}{B} &
        Time periods where hardness ratio is relatively larger, which is shown by the
        green parts of figure~\ref{hardness}. \\
        \addlinespace[2pt]
        \multirow{2}{*}{C} &
        Time periods where hardness ratio is relatively smaller, which is shown by the red
        parts of figure~\ref{hardness}. \\
        \bottomrule \bottomrule
    \end{tabular}
    \caption{\textbf{The definition of the time periods of the data set for the spectral and
    polarimetric analysis.}}
    \label{data}
\end{table*}

Three spectra of time period A may be inconsistent due to the systematic uncertainties of
instrumental responses among missions. We introduced a cross-calibration model,
\texttt{MBPO}, employed in the discovery paper to solve these discrepancies. It is defined as
$MBPO(E)=N\times(E/E_{br})^{\Delta\Gamma}$, where $\Delta\Gamma$ equals $\Delta\Gamma_1$ and
$\Delta\Gamma_2$ when $E <= E_{br}$ and $E > E_{br}$, respectively. 

\begin{figure}
    \centering
    \includegraphics[width=0.95\linewidth]{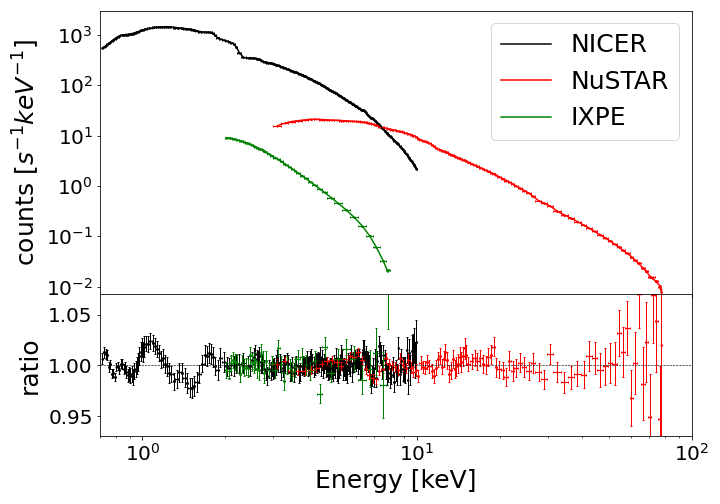}
    \caption{\textbf{Simultaneous fitting of \textit{IXPE}, \textit{NuSTAR}, and \textit{NICER} of
    time period A.} We fit time period A with the double Comptonization components model of
    \textit{IXPE} (green), \textit{NuSTAR} (red), and \textit{NICER} (black). The spectrum show high
    consistency with \texttt{MBPO} correction applied to them.
    {Alt text: One spectrum and data-to-model ratio plot of time period A.}}
    \label{TA}
\end{figure}

We fit the data of \textit{NICER}, \textit{NuSTAR}, and \textit{IXPE} of time period A
simultaneously with \texttt{MBPO} corrections applied to the two-Comptonization components model
given by equation (\ref{twocomp model}) to obtain the electron temperature of the Compton corona,
in which the \texttt{MBPO} parameters of \textit{NICER} are fixed to unity. The spectrum and
the data-to-model ratio are shown in figure~\ref{TA}, where the three spectra agree well with
each other. We note that \textit{IXPE} requires two \texttt{MBPO} corrections whose $E_{br}$
are set to 3 keV and 6 keV, respectively. We obtain an electron temperature $kT_e$ around
70 keV. In Section 3.3, we will analyze time periods B and C (where we use \textit{IXPE} and
\textit{NICER} data, but not \textit{NuSTAR} data) with $kT_e$ fixed at this value.

\subsection{Model Independent Analysis of PD}

\begin{figure}[h]
    \centering
    \subfigure{\includegraphics[width=0.45\textwidth]{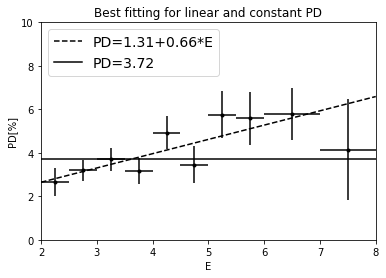}}
    \subfigure{\includegraphics[width=0.45\textwidth]{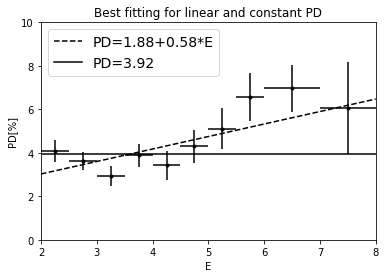}}
    \caption{\textbf{Fittings of time periods B and C with constant polarization model and
    linear polarization model.} The upper panel shows the best-fit results for time period B using
    the constant polarization model (solid line) and the linear polarization model (dashed line).
    The lower panel presents the corresponding fits for time period C, following the same scheme.
    See the main text for further details.
    {Alt text: Two plots for energy dependence of PD of time periods B and C.}}
    \label{energy dependency}
\end{figure}

We applied an event-by-event Stokes parameter approach to the \textit{IXPE} data, which
measures the Stokes parameters and gives weighting parameters for each individual event.
It allows for a more detailed analysis of the polarization compared to binning data in
scatting/emission angles (\cite{Kislat2015}). We extracted the Stokes parameters in four
energy bins (2-3 keV, 3-4 keV, 4-6 keV, and 6-8 keV) of time periods B and C, and calculated
the PD in a spectral-model independent way by using \texttt{PCUBE} implemented in
\texttt{ixpeobssim} with the response
matrices\footnote{\texttt{ixpe\_obssim20211209\_alpha075\_v013}, where 20211209 means the 
date that the response is available and lasting for 6 months, alpha075 means that the response
files take event weighting into account, and v013 is the version of the response file} that
takes event weighting into account. We observe an apparent trend of the energy dependence
in PD during both time periods B and C (the relatively hard and soft parts of observation 1).
We also analyzed observation 2 in the same way and obtained insignificant energy dependence of
PD, probably due to low photon statistics. Hereafter, we focus on time periods B and C and
tested a constant and a linear polarization model. The fitting results are
presented in figure~\ref{energy dependency}.

The constant polarization model yields best-fit PDs of $3.62\%$ and $3.56\%$ for time periods
B and C, respectively. The $\chi^2$ values are 14.73 and 28.06 with 9 degrees of freedom
(\textit{d.o.f}) for time periods B and C, respectively. These results indicate that the
constant PD model is rejected at $90.0\%$ and $99.9\%$ confidence levels for time periods B
and C, respectively. The apparent increase in PD with energy might be modeled by a linear
polarization model, $PD = a_0 + a_1 \times E$. The model yields the best-fit values of
$a_0 = 1.50 \pm 0.77$ and $a_1 = 0.65 \pm 0.20$ for the time period B with a $\chi^2$ of 3.53
for 8 \textit{d.o.f}, and $a_0 = 2.02\pm 0.67$ and $a_1 = 0.54 \pm 0.18$ for time period C
with a $\chi^2$ of 11.89 for the same \textit{d.o.f}, respectively. The linear model provides
a significant improvement in the fitting. We conclude that there is a clear positive energy
dependence of PD in both the hard and soft parts of observation 1. These
results are in good agreement with those reported in the discovery paper. In addition, we may
also conclude that time periods B and C exhibit a similar energy dependence in PD regardless
of the different hardness ratio.

\subsection{Spectrum and Polarimetric Analysis}

To investigate the source spectrum in detail, we analyzed \textit{NICER} and \textit{IXPE}
data for time periods B and C. Here we replaced the soft Comptonization component model,
\texttt{NthComp}, with \texttt{thComp}, a more physical model that agrees much better with
the actual Monte Carlo spectra from Comptonization (\cite{Zdziarski2020}). The model has no
normalization parameter since it follows the seed photons' normalization. In other words,
\texttt{thComp} is a model that considers both direct emissions from the thermal disk and the 
Comptonized emissions of the corona. Then our model formula is represented by,

\begin{equation}
    \centering
    \mathrm{MBPO} * \mathrm{MBPO} * \mathrm{TBABS} *
    (\mathrm{thComp}*\mathrm{diskBB} + 
    \mathrm{NthComp} + \mathrm{relxillCp} + \mathrm{xillverCp}).
    \label{physical model}
\end{equation}

In the following spectrum and polarimetric analysis, we will use the model of
equation (\ref{physical model}) to fit time periods B and C. Since \textit{NuSTAR} observed 
Cyg X-1 for a short time, we decided to perform the spectrum and polarimetric analysis
using data from \textit{NICER} and \textit{IXPE} only but, where
\textit{NICER} data provide the information for the spectrum, and \textit{IXPE} data
give the polarization properties.

We freed the absorption during the fit to represent a possible change in absorption on
average. The electron temperature $kT_e$ was fixed at 70 keV obtained by using \textit{NuSTAR}
data (Section 3.1). To determine the normalization of the \texttt{diskBB}, we refer to the
\textit{NICER} data of period 0 (figure~\ref{LCnicerixpe}); it does not suffer from severe
absorption and has the best photon statistics among time periods. We obtained the normalization
to be $1.72 \times 10^6$, and fixed it for time periods B and C. This implies that the inner
disk radius remains unchanged. The seed photon temperature $kT_{bb}$ and electron
temperature $kT_e$ of the \texttt{NthComp} are tied to the inner disk temperature $kT_{in}$
of \texttt{diskBB} and $kT_e$ of \texttt{ThComp} (fixed at 70 keV), respectively. The
inner disk radius $R_{in}$ of the relativistic reflection component \texttt{RelxillCp} is free
to vary to absorb small structures around the iron K line around 6.4 keV, and the photon index
$\Gamma$ is tied to that of the hard Compton component in \texttt{NthComp}. We fixed the other
parameters to be the same as in the discovery paper, but freed the normalizations. The standard
reflection component \texttt{xillverCp} is connected to the relativistic reflection component.
We allowed \texttt{MBPO} parameters to differ from what we obtained in time period A, because
Cyg X-1 is very bright and may cause gain shifts during observation.

\begin{figure*}[h]
    \centering
    \subfigure{\includegraphics[width=0.49\textwidth]{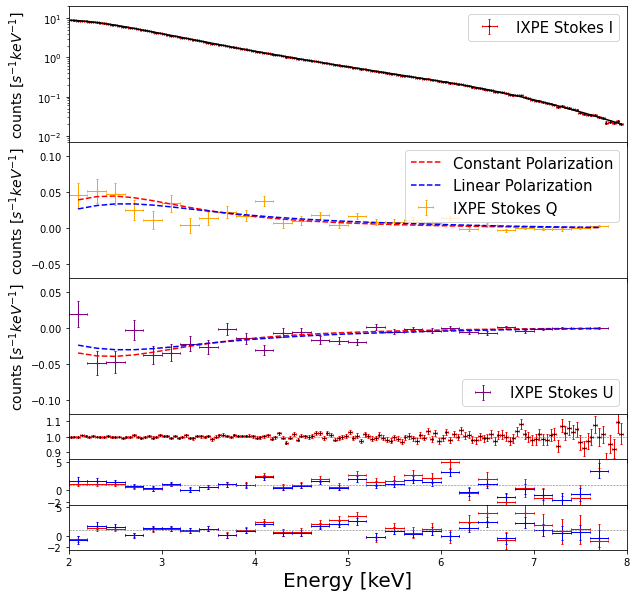}}
    \subfigure{\includegraphics[width=0.49\textwidth]{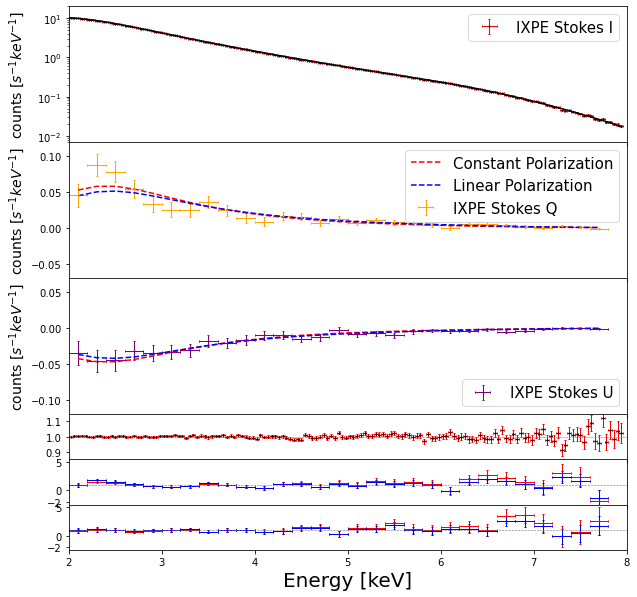}}
    \caption{\textbf{Spectrum and data-to-model ratio of time periods B and C with the \texttt{MBPO}
    correction and polarization models applied.} The left panel illustrates the data-to-model ratio
    of time period B with constant and linear polarization models applied to the data, respectively.
    The upper 3 panels show spectra of Stokes I (black), Q (orange), and U (purple), and the best
    fittings of constant and linear polarization models are given in red and blue dashed lines,
    respectively. The lower 3 panels exhibit the data-to-model ratio with the same color. The right
    panel shows those in the same color scheme.
    {Alt text: Two spectrum and data-to-model ratio plots of time periods B and C.}}
    \label{newfitting}
\end{figure*}

The constant and linear polarization models were then applied to only the hard Comptonization
component that dominates the spectrum in the 2 -- 8 keV \textit{IXPE} band. The fitting results
are summarized in table~\ref{newfit}, and the spectra and the data-to-model ratio are shown in
figure~\ref{newfitting}. The constant polarization model gives a total $\chi^2$ of 300.64
with a \textit{d.o.f} of 205, while the linear polarization model gives 289.01/204 for the
time period B. Meanwhile, $\chi^2/d.o.f$ for the time period C are 159.90/205 and 155.41/204
for constant and linear polarizations, respectively. The F-test gives a probability greater
than $99\%$ and $98\%$, suggesting that the fit improves significantly with the linear
polarization model; this result agrees with the model-independent analysis presented
in Section 3.1 and aligns with the findings of the discovery paper.

\begin{figure}[h]
    \centering
    \includegraphics[width=0.95\linewidth]{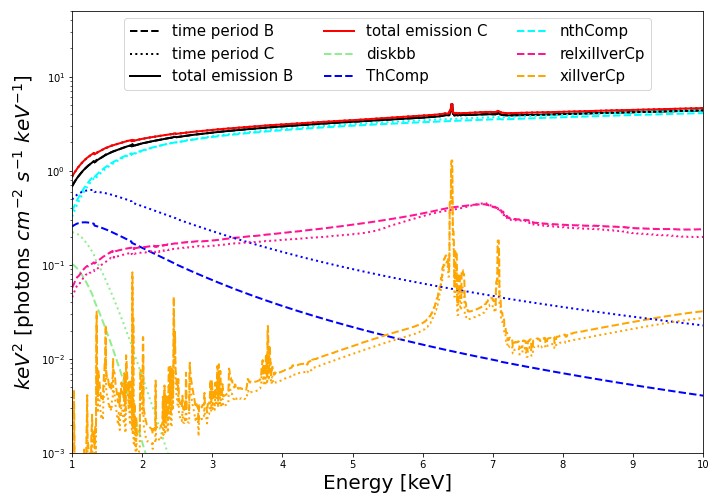}
    \caption{\textbf{Best fitting individual components for time periods B and C.} The black and red
    solid lines are the total emissions of time periods B and C. The best fitting individual
    components are labeled as dashed and dotted lines for time periods B and C with different
    colors, where the disk emission is shown in light green, the soft Comptonization component is
    shown in blue, the hard Comptonization component is shown in cyan, and the relativistic and
    standard reflection components are shown in deep pink and orange.
    {Alt text: A plot showing model contributions for time periods B and C.}}
    \label{components}
\end{figure}

For the spectral parameters, we observed an inner disk temperature $kT_{in} \sim 0.15$ keV,
which is $\sim 2$ times smaller than that presented in the discovery paper ($kT_{in} = 0.319$
keV). This will help us explain the high PD with a canonical inclination angle since seed
photons with a lower temperature have to experience more times of scattering to reach the
\textit{IXPE} energy band, leading to a higher PD. The photon indices in our results are
consistent with those reported in the discovery paper, while the inferred inner disk radius
appears slightly larger. This difference may reflect subtle variations in the modeling of the
reflection component, such as the treatment of the broad iron line (\cite{Ding2024}). The
differences between time periods B and C are mainly caused by different intrinsic spectrum,
but not different absorption, and the individual components of them are shown in 
figure~\ref{components}. We find
that, although the disk component in the period C increases by a factor of two, and leads to an
increase of the soft emission, the inner disk temperature remains stable. We thus conclude that
the seed photon temperature is rather low ($kT_{in} \sim 0.15$ keV) throughout observation 1.


\small
\begin{table*}
\centering
\begin{tabular}{lllcccc}
    \hline\hline
    \multirow{2}*{Model} & \multirow{2}*{Parameter (unit)} & \multirow{2}*{Description} 
                         & \multicolumn{4}{c}{Time Periods}\\
      &  & & \multicolumn{2}{c}{B} & \multicolumn{2}{c}{C} \\
    \toprule
    \hline



    
    $\texttt{TB}_{\texttt{ABS}}$ & $N_H$ ($10^{22}$ $\mathrm{cm}^{-2}$) & Hydrogen column density &
                        \multicolumn{2}{c}{$0.569_{-0.006}^{+0.004}$} & 
                        \multicolumn{2}{c}{$0.608_{-0.005}^{+0.008}$} \\
    \hline
    \multirow{2}*{\texttt{diskBB}}   & $kT_{in}$ (keV) & Peak disk temperature &
                        \multicolumn{2}{c}{$0.133_{-0.0015}^{+0.0015}$} & 
                        \multicolumn{2}{c}{$0.147_{-0.0010}^{+0.0013}$} \\
                                     &   N          & Normalization & 
                        \multicolumn{2}{c}{$1.72\times10^6$ (fix)} &
                        \multicolumn{2}{c}{$1.72\times10^6$ (fix)} \\
    \hline
    \multirow{2}*{\texttt{ThComp}}   & $\Gamma$     & Photon index &
                        \multicolumn{2}{c}{$4.46_{-0.15}^{+0.15}$} & 
                        \multicolumn{2}{c}{$4.03_{-0.06}^{+0.16}$} \\
                                     & $kT_e$ (keV) & Electron temperature &
                        \multicolumn{2}{c}{70 (fix)} & 
                        \multicolumn{2}{c}{70 (fix)} \\
    \hline
    \multirow{6}*{\texttt{Polarization}} &  \multirow{2}*{PD (\%)} &
                                           \multirow{2}*{Polarization degree} &
                        \multicolumn{2}{c}{$3.58_{-0.45}^{+0.45}$} & %
                        \multicolumn{2}{c}{$4.27_{-0.39}^{+0.39}$} \\ %
                  & & & \multicolumn{2}{c}{[$1.35_{-1.13}^{+1.13}$]} &
                        \multicolumn{2}{c}{[$3.07_{-1.01}^{+1.01}$]} \\
                                     & PD slope (\%) & Slope of linear polarization &
                        \multicolumn{2}{c}{--- \ [$0.87_{-0.40}^{+0.40}$]} & 
                        \multicolumn{2}{c}{--- \ [$0.47_{-0.37}^{+0.37}$]}\\
                                     & \multirow{2}*{PA (degree)} & 
                                       \multirow{2}*{Polarization angle} &
                       \multicolumn{2}{c}{$-20.92_{-3.55}^{+3.52}$} & %
                       \multicolumn{2}{c}{$-19.48_{-2.61}^{+2.61}$} \\ %
                 & & & \multicolumn{2}{c}{[$-21.12_{-3.38}^{+3.42}$]} &
                       \multicolumn{2}{c}{[$-19.70_{-2.60}^{+2.60}$]} \\
                                     & PA slope (degree) & Slope of linear polarization &
                        \multicolumn{2}{c}{--- \ [0 (fix)]} & 
                        \multicolumn{2}{c}{--- \ [0 (fix)]} \\
    \hline
    \multirow{3}*{\texttt{NthComp}}  & $\Gamma$    & Photon index &
                        \multicolumn{2}{c}{$1.62_{-0.010}^{+0.009}$} & 
                        \multicolumn{2}{c}{$1.59_{-0.012}^{+0.013}$} \\
                                     & $kT_e$ (keV) & Electron temperature &
                        \multicolumn{2}{c}{70 (fix)} & 
                        \multicolumn{2}{c}{70 (fix)} \\
                                     & norm        & Normalization &
                        \multicolumn{2}{c}{$1.629_{-0.057}^{+0.054}$} & 
                        \multicolumn{2}{c}{$1.552_{-0.037}^{+0.042}$} \\
    \hline
    \multirow{4}*{\texttt{RelxillCp}} & $R_{in}$ ($r_g$) & Inner disk radius &
                        \multicolumn{2}{c}{$6.49_{-0.93}^{+2.61}$} & 
                        \multicolumn{2}{c}{$24.30_{-6.93}^{+6.56}$} \\
                                      & $\mathrm{log}_{10}(\xi/[\mathrm{erg\ cm} \ \mathrm{s}^{-1}])$ &
                                      Ionization parameter &
                        \multicolumn{2}{c}{$3.11_{-0.056}^{+0.052}$} & 
                        \multicolumn{2}{c}{$3.19_{-0.095}^{+0.083}$} \\
                                     & $kT_e$ (keV) & Electron temperature &
                        \multicolumn{2}{c}{140 (fix)} & 
                        \multicolumn{2}{c}{140 (fix)} \\ 
                                     & norm ($10^{-2}$) & Normalization &
                        \multicolumn{2}{c}{$5.59_{-0.49}^{+0.52}$} & 
                        \multicolumn{2}{c}{$5.57_{-0.56}^{+0.60}$} \\
    \hline
    \multirow{4}*{$\texttt{MBPO}_{\texttt{IXPE}}$} & $\Delta \Gamma_1$ ($10^{-2}$) 
                                                   & low-energy power-law index &
                        \multicolumn{2}{c}{$-37.60_{-1.76}^{+1.68}$} & 
                        \multicolumn{2}{c}{$-67.87_{-2.82}^{+2.75}$} \\
                                 & $\Delta \Gamma_2$ ($10^{-2}$) & high-energy power-law index &
                        \multicolumn{2}{c}{$-16.60_{-1.30}^{+1.23}$} & 
                        \multicolumn{2}{c}{$-29.29_{-2.24}^{+2.22}$} \\
                                 & $E_{br}$ (keV) & Break energy &
                        \multicolumn{2}{c}{3 (fix)} & 
                        \multicolumn{2}{c}{3 (fix)} \\
                                 & N & Normalization &
                        \multicolumn{2}{c}{$0.946_{-0.005}^{+0.006}$} & 
                        \multicolumn{2}{c}{$0.959_{-0.016}^{+0.018}$} \\
    \hline
    \multirow{2}*{$\texttt{MBPO}_{\texttt{IXPE}}$} & $\Delta \Gamma_2$ ($10^{-2}$) 
                                                   & high-energy power-law index &
                        \multicolumn{2}{c}{$-32.38_{-9.23}^{+9.19}$} & 
                        \multicolumn{2}{c}{$-65.27_{-8.72}^{+8.76}$} \\
                                 & $E_{br}$ (keV) & Break energy &
                        \multicolumn{2}{c}{6 (fix)} & 
                        \multicolumn{2}{c}{6 (fix)} \\
    \hline 
    \multicolumn{2}{c}{$\tau_S$} & Optical depth of the soft part &
                        \multicolumn{2}{c}{0.297} &
                        \multicolumn{2}{c}{0.357} \\
    \hline 
    \multicolumn{2}{c}{$\tau_H$} & Optical depth of the hard part &
                        \multicolumn{2}{c}{0.969} &
                        \multicolumn{2}{c}{1.017} \\
    \hline
    \bottomrule
    \end{tabular}
    \caption{\textbf{Best fitting parameters of time period B and C with two-Comptonization 
components model with constant and linear polarization models.} We applied a constant and
linear polarization model. The results of the latter are shown in a blanket
(e.g., [$-19.70_{-2.60}^{+2.60}$]).} 
    \label{newfit} 
\end{table*}

The optical depth $\tau$ of the coronae during periods B and C can be estimated using the
obtained spectral parameters. Here we assume simple corona geometries, namely a spherical
geometry and a slab geometry, where the optical depth $\tau$ of a spherical geometry can be
evaluated by equation (\ref{spectrum index1}) (e.g., \cite{Zdziarski1996}), 
\begin{center}
\begin{equation}
    \alpha = [\frac{9}{4}+\frac{1}{(kT_{e}/m_ec^2)\tau(1+\tau/3)}]^{1/2}-\frac{3}{2}, 
\label{spectrum index1}
\end{equation}
\end{center}
and that of a slab geometry is evaluated by equation (\ref{spectrum index2}) (\cite{Pozdnyakov1983}),
\begin{equation}
    \centering
    \alpha = [\frac{9}{4}+\frac{\pi^2}{12}\frac{m_ec^2}{kT_{e}(\tau+\frac{2}{3})^2}]^{1/2}-
    \frac{3}{2}.
    \label{spectrum index2}
\end{equation}

We assumed a slab geometry for the hard Comptonization component and a spherical geometry
for the soft Comptonization component, and obtained the $\tau$ as listed in table~\ref{newfit}.
The optical depths of the hard component are compatible with that presented in the discovery paper.
The optical depths of both the hard and soft components are similar to those reported by
\citet{Makishima2008} using the observation of Cyg X-1 in 2005. Therefore, the
disk/corona in 2022 hard state is in a typical condition of Cyg X-1. As described in Section 1,
different corona geometries can produce a similar spectrum and are not distinguishable, and X-ray
polarization will solve this degeneracy. In the next section, we will investigate the polarization
properties in detail by comparing with a dedicated simulation.

\section{Discussion of the Corona Geometry}

In Section 3, we employed the two-Comptonization components model and succeeded in reproducing
the spectral changes with reasonable model parameters. Here, we aim to reveal the corona
geometry using the spectrum and polarization properties obtained. The obtained inner disk temperature
\textcolor{black}{(0.15 keV) is approximately} half of the value reported in the discovery paper
\textcolor{black}{(0.3 keV)}; this would require a larger number of scatterings for seed photons to
reach the \textit{IXPE} band (2 -- 8 keV) and produce larger PD. To examine this expectation
quantitatively, \textcolor{black}{we simulated the spectrum and polarization in a single-temperature
corona model to mimic the hard corona that is relevant to the polarization properties in the 
\textit{IXPE} band with two types of corona geometries}, a \textcolor{black}{sandwiching} slab corona
and a wedge-shaped corona with a truncated disk. \textcolor{black}{The simulation is based on the ray
tracing code MONK\footnote{https://projects.asu.cas.cz/zhang/monk} (\cite{Zhang2019}), a Monte Carlo
radiative transfer code dedicated to calculate the Comptonization spectrum with a Klein-Nishina cross
section and relativistic effects taken into account, which was also used in the discovery paper}.
\textcolor{black}{MONK simulates disk radiation accounting for relativistic effects (i.e. gravitational
and motion frequency shifts), however, we should be reminded that it assumes a single-temperature
black body (BB) radiation, which is different from the emission from a standard disk.}

We first simulate the spectrum and investigate the polarization properties of
$kT=0.15$ keV, and compare them with the case of $kT=0.3$ keV later. We tuned the accretion
rate in each simulation so that the disk temperature settles at approximately $0.15$ keV. We also
tuned the optical depth for both coronae so that the photon indices are similar to the observed value
($\sim 1.7$). For \textcolor{black}{the} wedge-shaped corona, the inner edge of the corona is at the
ISCO and the outer edge is at 40 $r_g$, where the disk is truncated. The opening angle is $10^{\circ}$
and the electron temperature is assumed to be 100 keV, the same as in the discovery paper. The slab
corona extends from the ISCO to 100 $r_g$ with the same electron temperature, and the half height of
the corona is set at 2 $r_g$. \textcolor{black}{We do not account for the reflection of the
Compton-scattered photons at the disk and the returning
radiation in our simulation as they are not fully implemented in MONK yet.} The parameters for
simulation are summarized in table~\ref{simp}.

\begin{table*}[htb]
    \centering
    \begin{tabular}{llcccc}
    \hline
    \toprule
    \multirow{2}*{Parameters} & \multirow{2}*{Description} & 
    \multicolumn{2}{c}{Value [unit]} \\
      &   &   Sandwiching Slab Corona & Wedge-shaped Corona \\
    \hline
    $\iota$ & \textcolor{black}{inclination angle} & \multicolumn{2}{c}{$60^{\circ}$}\\
    $a$ & BH spin parameter & $a=0\ /a=0.9$ & $a=0\ /a=0.9$ \\
    $R_{in}$ & inner disk radius & ISCO & 40 $r_g$ \\
    $C_{in}$ & inner corona radius & ISCO & ISCO \\
    $C_{out}$ & outer corona radius & 100 $r_g$ & 40 $r_g$ \\
    $h$ & corona half height & 2 $r_g$  & --  \\
    $\theta$ & opening angle of the corona & -- & $10^{\circ}$ \\
    $\dot{m}$ & accretion rate & 0.01/ 0.001 & 0.02 /0.02 \\
    $kT_{bb}$ & \textcolor{black}{disk temperature} & \multicolumn{2}{c}{$\sim0.15$ keV} \\
    $kT_e$ & electron temperature & 100 keV & 100 keV \\
    $\tau$ & optical depth & 0.5 & 2 \\   
    \bottomrule
    \end{tabular}
    \caption{\textbf{Parameters for spectrum and polarimetric simulations \textcolor{black}{used for
    figure~\ref{sim-spec} and figure~\ref{sim-pol}. For figure~\ref{sim-compare}, we used the
    parameters in the left column (sandwiching slab corona) but increase the accretion rate by a
    factor of 20 to simulate the spectrum and polarization properties for disk temperature 0.3 keV
    case. We take neither the reflection of the Compton-scattered photons nor the returning
    radiation into account in the simulation.}}}
    \label{simp}
\end{table*}

\begin{figure*}[h]
    \centering
    \subfigure{\includegraphics[width=0.45\textwidth]{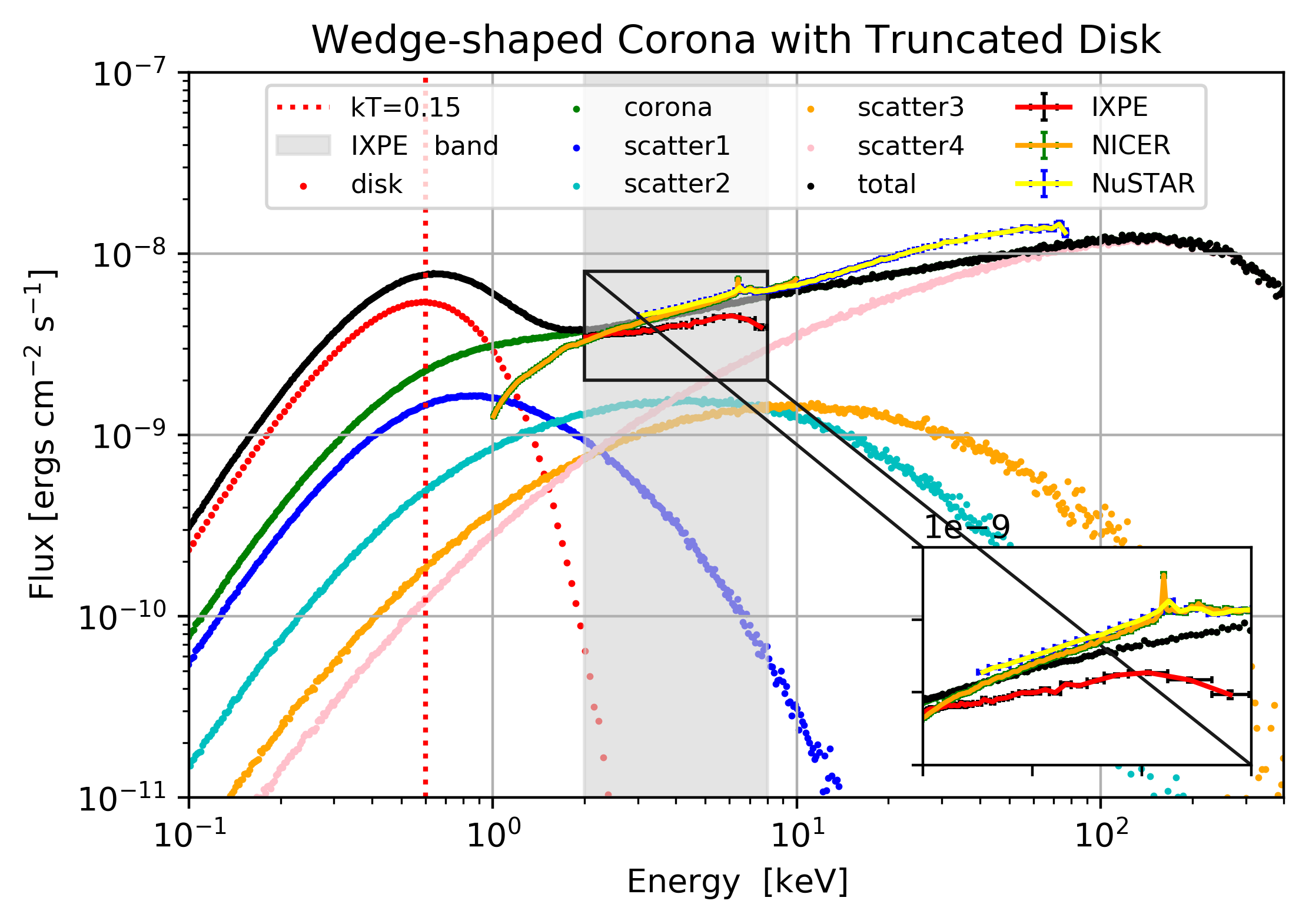}}
    \subfigure{\includegraphics[width=0.45\textwidth]{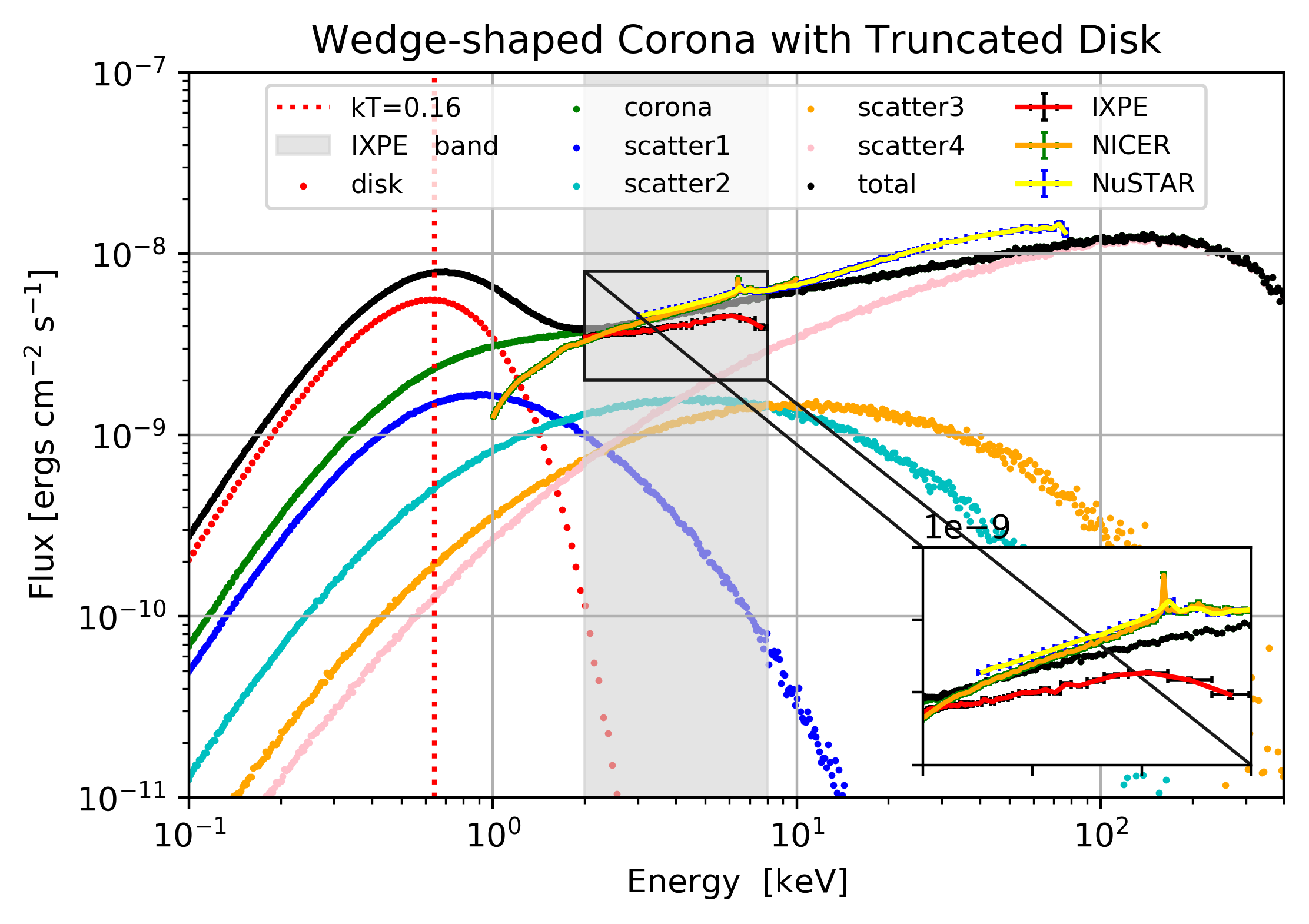}}
    \subfigure{\includegraphics[width=0.45\textwidth]{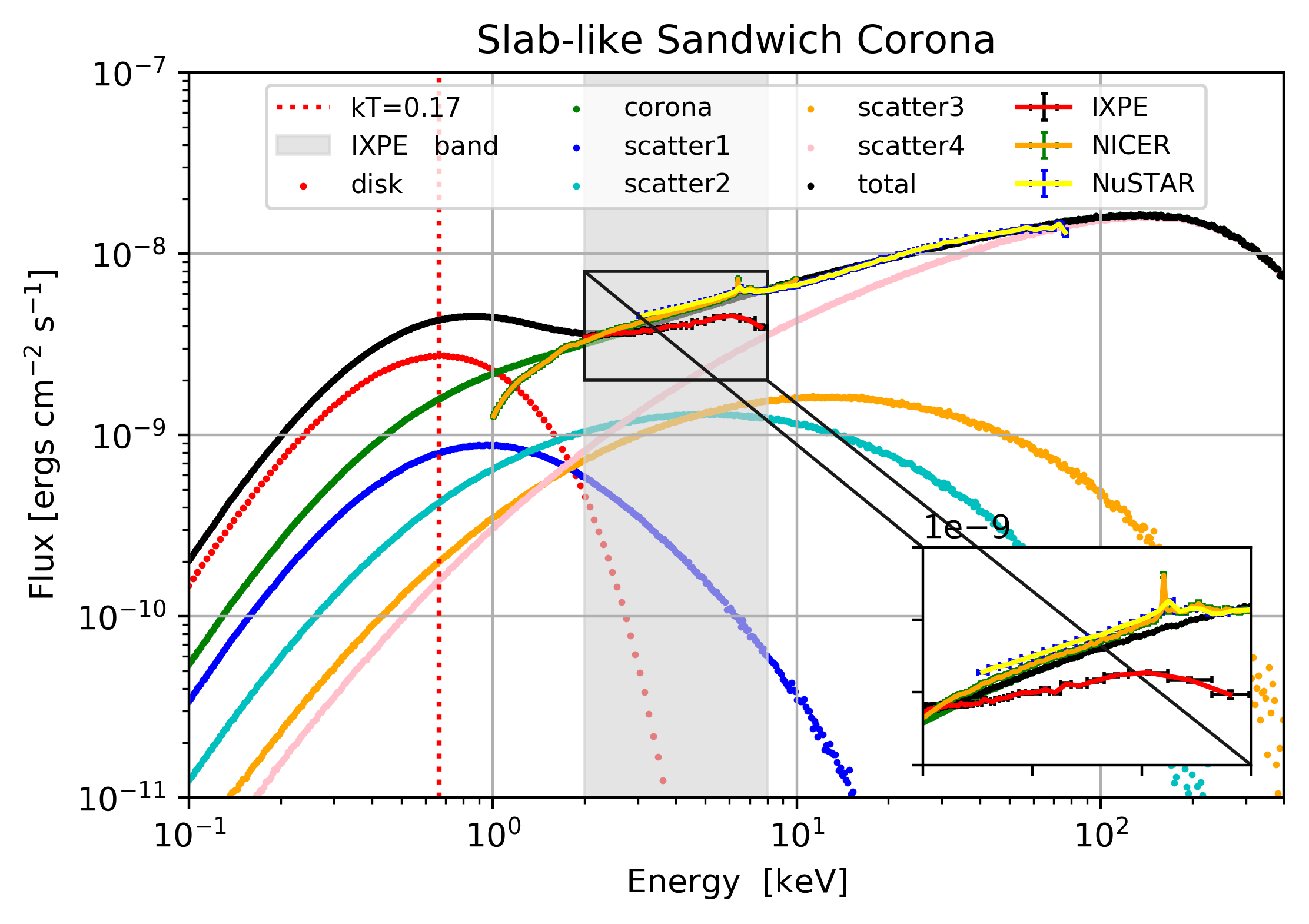}}
    \subfigure{\includegraphics[width=0.45\textwidth]{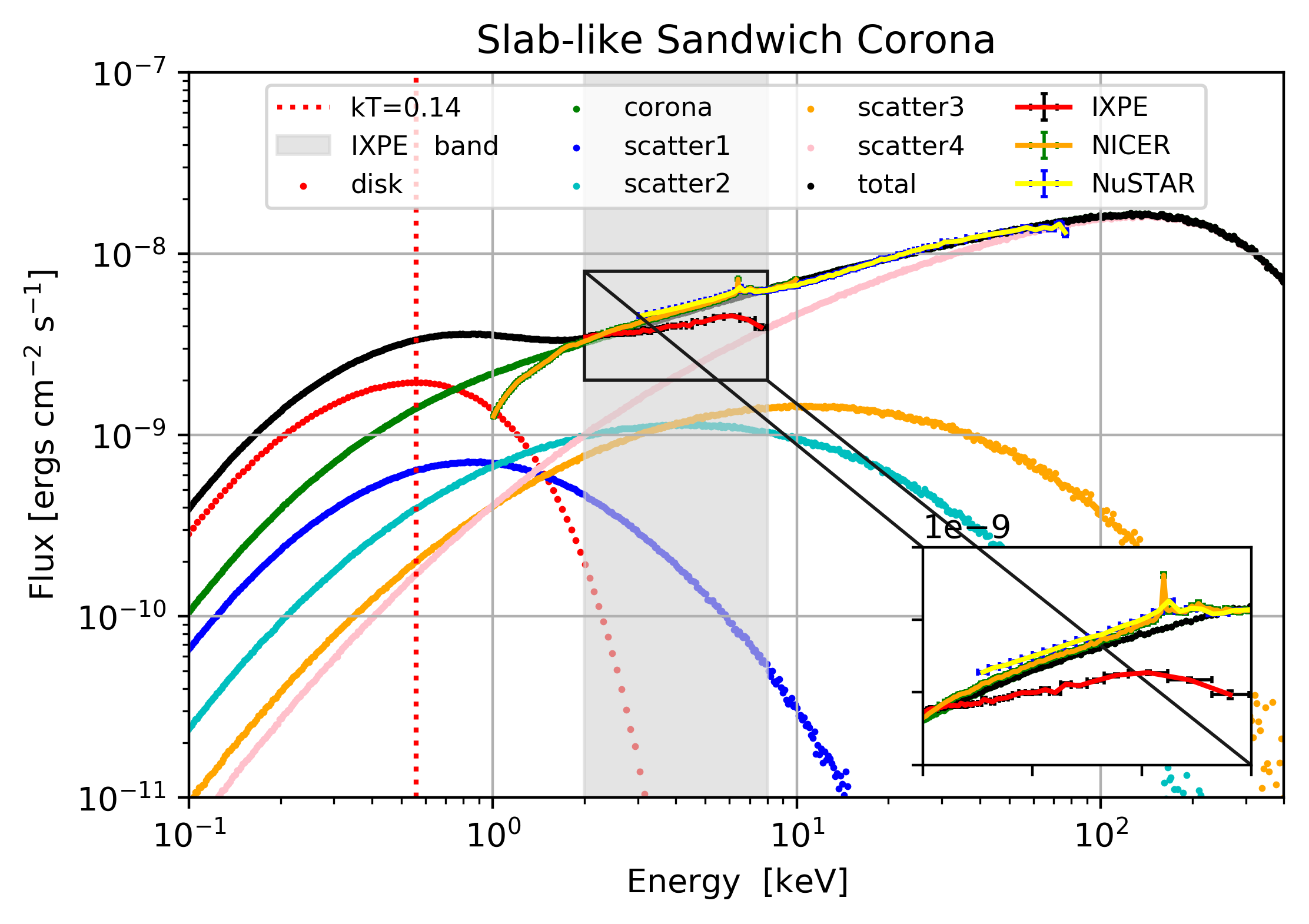}}
    \caption{\textbf{The simulated spectrum with four model configurations and the simultaneous
    observed \textit{IXPE}, \textit{NICER}, and \textit{NICER} data}.} The upper panel shows the
    spectra of wedge-shaped corona with truncated disk, while the lower panel illustrates
    those of the sandwiching slab corona, both Schwarzschild (left) and Kerr (right) BH cases with
    $a = 0.9$, respectively. The total spectrum is shown in black, the disk emission is shown in red,
    and the total corona emission is shown in green. Blue, cyan, orange, and pink stand for
    spectra of the photons that scattered 1, 2, 3, and at least 4 times. \textcolor{black}{The
    parameters for the simulation are listed in table~\ref{simp} and the inclination is assumed to
    be $60^{\circ}$ to investigate the properties in detail.} The gray area represent the
    \textit{IXPE} energy band, and the simultaneous observation data of \textit{IXPE},\textit{NICER},
    and \textit{NuSTAR} are shown in red, orange, and yellow points with error bars, respectively.
    {Alt text: Four simulated spectrum \textcolor{black}{and data} plots for hard state of Cyg X-1.}
    \label{sim-spec}
\end{figure*}

\textcolor{black}{Figure~\ref{sim-spec}} summarizes the simulated spectra \textcolor{black}{with
$kT=0.15$ keV of four configurations, the sandwiching slab corona and the wedge-shaped corona with
non-spinning BH and spinning BH without reflection. We also demonstrate the spectra of different
numbers of scatterings and the simultaneous observed data of \textit{IXPE}, \textit{NICER}, and
\textit{NuSTAR}. We should be aware that we do not account for absorption in our simulation
as it has large uncertainties.} In the wedge-shaped corona, photons that \textcolor{black}{were}
scattered \textcolor{black}{twice} dominate the low-energy part of the spectrum in the \textit{IXPE}
band, whereas photons that were scattered at least four times dominate the high-energy
part. In contrast, photons that were scattered at least four times dominate the entire \textit{IXPE}
energy band in the sandwiching slab corona geometry. The subplots
show the comparison between the observed and simulated spectra in the \textit{IXPE} energy band in
detail. The simulation can approximately reproduce the \textit{NICER} data in the \textit{IXPE} energy
band, but some discrepancies appear below 2 keV, where the observed spectrum shows stronger curvature
primarily due to absorption. As the absorption does not largely affect the spectrum in the 2 -- 8 keV
band, this discrepancy will not significantly affect the subsequent polarization analysis. Therefore,
we safely ignore the correction for absorption.

\begin{figure*}[h]
    \centering
    \subfigure{\includegraphics[width=0.45\textwidth]{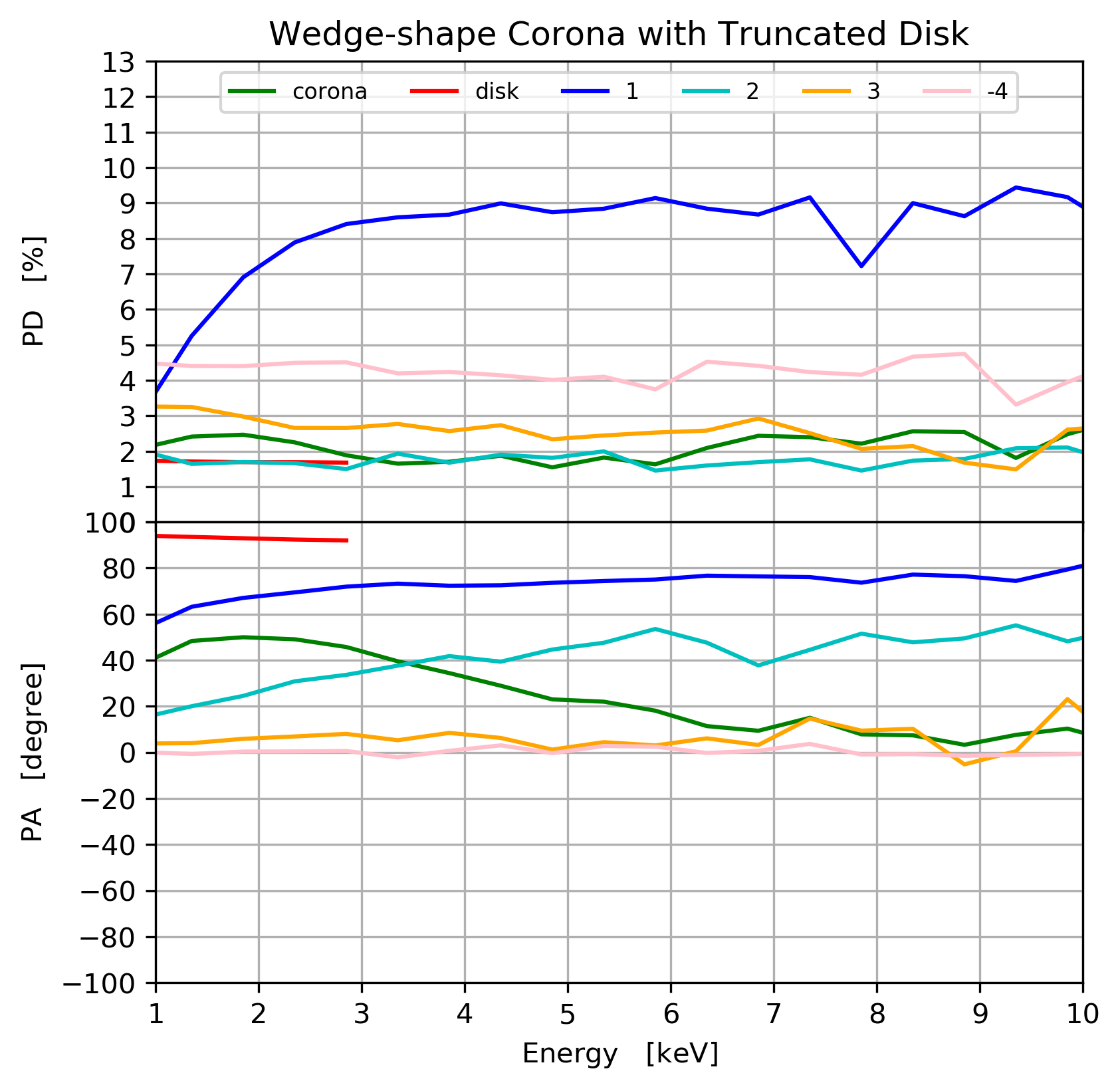}}
    \subfigure{\includegraphics[width=0.45\textwidth]{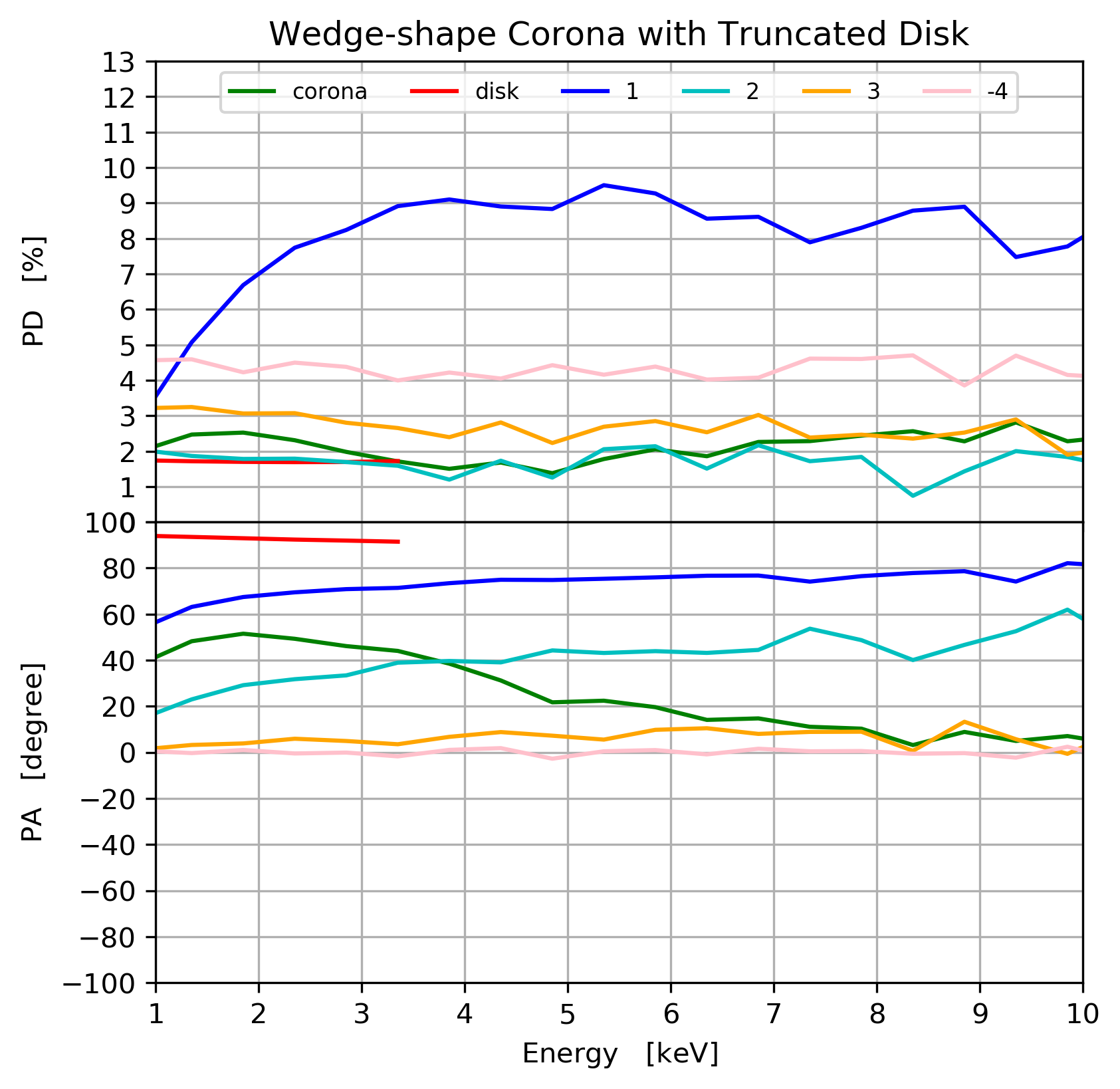}}
    \subfigure{\includegraphics[width=0.45\textwidth]{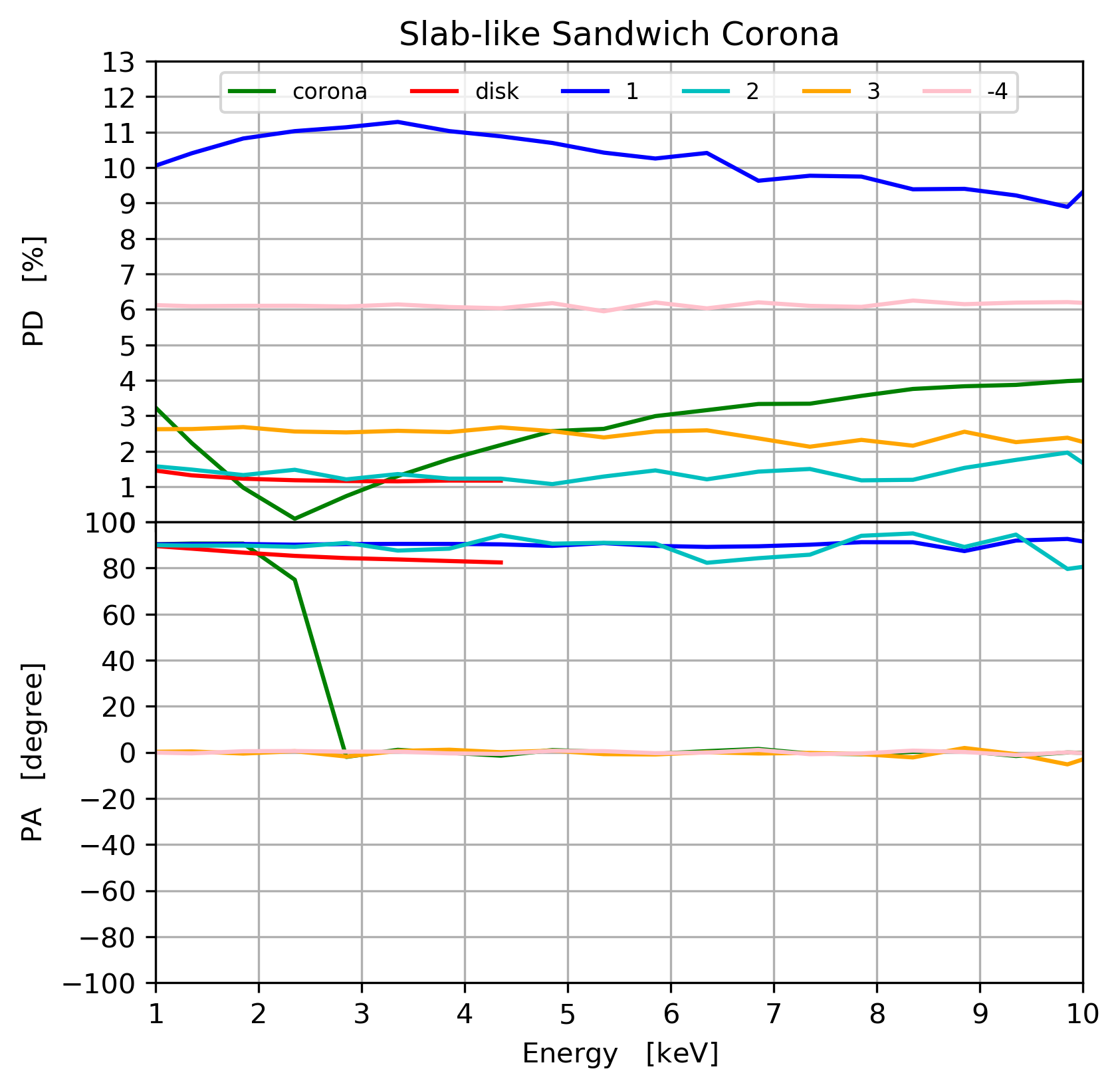}}
    \subfigure{\includegraphics[width=0.45\textwidth]{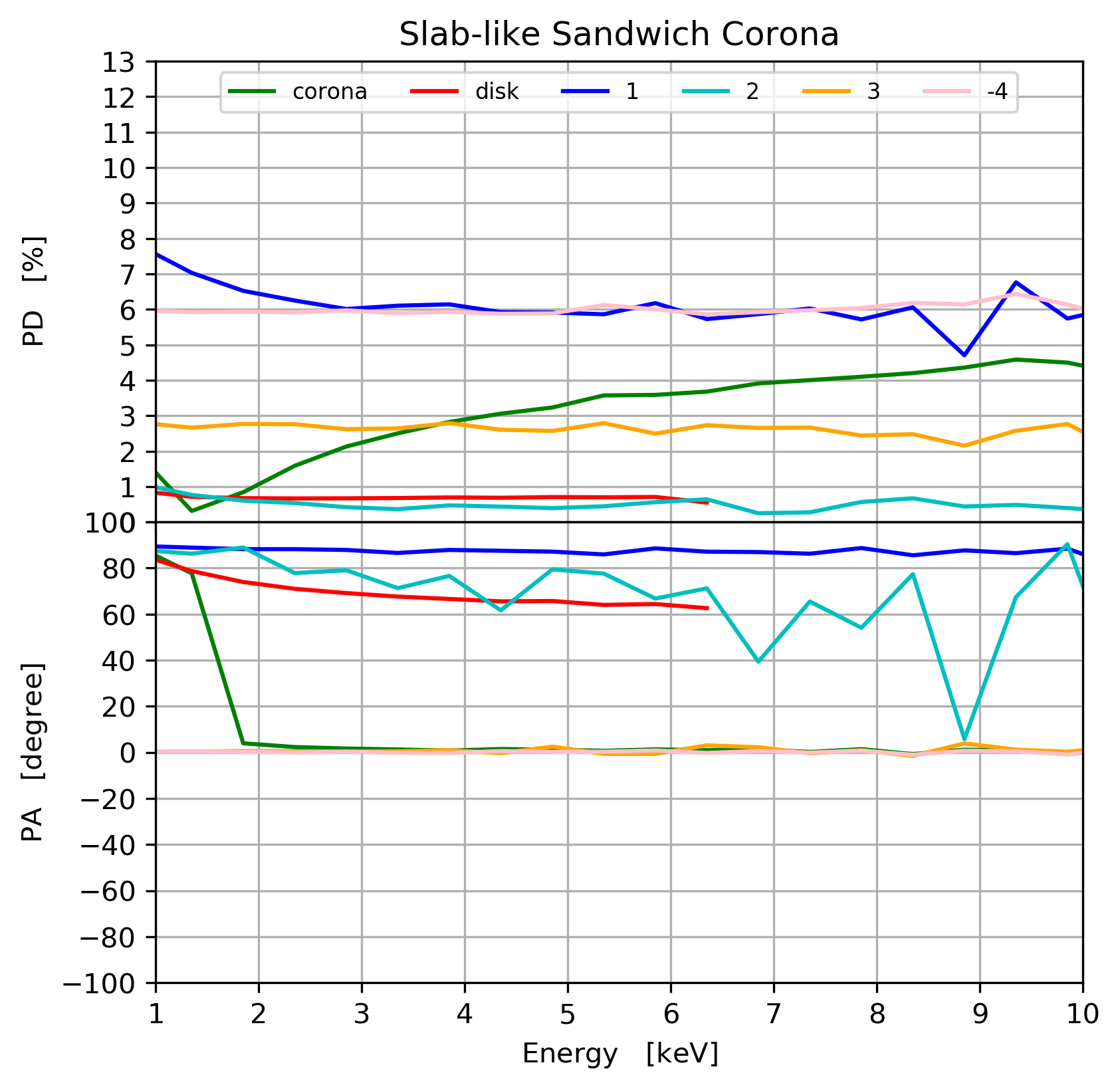}}
    \caption{\textbf{The simulated polarization with four model configurations.} The upper
    panel shows the PD and PA of wedge-shaped corona with truncated disk geometry, while the lower
    panel illustrates those of the sandwich slab corona geometry, both Schwarzschild (left)
    and Kerr (right) BH case with $a = 0.9$, respectively. The polarization is perpendicular
    to the disk if the PA is $0^{\circ}$. \textcolor{black}{We used the same parameters for the
    simulation} and adopt the same color scheme as in figure~\ref{sim-spec} for consistency. The
    polarization of the total spectrum is omitted, as it does not significantly affect the results.
    {Alt text: Four simulated polarizations plots for hard state of Cyg X-1.}}
    \label{sim-pol}
\end{figure*}

We then examine the polarization properties of the four model configurations. Figure~\ref{sim-pol}
shows the energy dependence of PD and PA. Here we show the results for the inclination angle of
$60^{\circ}$ to investigate the properties in detail (higher inclination angle gives larger PD),
but the trends are similar to those of the $30^{\circ}$ case. The figure shows common trends
among four configurations. Firstly, the photons that were scattered once give the highest PD,
and the PA is parallel to the disk. For multiple scattering photons, the PD increases with the
number of scatterings, and the PA is always perpendicular to the disk when the photon is
scattered at least 3 times. 

Then, we will investigate the polarization properties of the wedge-shaped corona geometry
\textcolor{black}{and the sandwiching slab corona geometry} in more detail. \textcolor{black}{In the
wedge-shaped corona geometry,} we suggest that most of the \textcolor{black}{scattered} photons come
from the inner part of the disk (closest to the corona). \textcolor{black}{Among them, photons} entering
the corona vertically have more chances \textcolor{black}{of reaching} us with a single scattering.
This is why the PA of the \textcolor{black}{singly} scattered photons is almost parallel to the disk.
With an increasing number of \textcolor{black}{scatterings, the} seed photons will pass through the
corona horizontally, giving the PA almost perpendicular to the disk. The total PD is almost constant
($\sim 2.5\%$) in the \textit{IXPE} band, \textcolor{black}{while} the PA exhibits a clear energy
dependence. We observed that the PA is larger, or deviates from the normal of the disk, in low-energy.
In \textcolor{black}{the \textit{IXPE} 2--8 keV energy band, the one-time and two-time scattered photons
dominate the low-energy part of the spectrum while the photons that were scattered at least four times
dominate the high-energy parts. This is why we observe the energy dependence of PA.}

\textcolor{black}{In the case of sandwiching slab corona}, we observe that the seed photons behave in
a slightly different way. The photons that were scattered once pass through the corona vertically,
resulting in a constant PA of $90^{\circ}$. However, photons that were scattered at least three times
pass through the disk almost horizontally, leading to a constant PA of $0^{\circ}$ (perpendicular to
the disk).  \textcolor{black}{See the lower panel of figure~\ref{sim-spec}. Since the PA are different
by $90^{\circ}$, the single or double scattered photons and the photons that were scattered more than
three times
depolarize each other, resulting in a decrease in PD in the low-energy band. As photons
that were scattered at least four times start to dominate the spectrum in the high-energy band, we
observed an increase in PD in the high-energy band, presenting
an apparent energy dependence}. Interestingly, the observed value and energy dependence of PD
of the time periods B and C (figure~\ref{energy dependency}) agree well with those of the slab
corona. In contrast, the wedge-shaped corona with a truncated disk shows a somewhat different
energy dependence of polarization. The data of Cyg X-1 observed in 2022 May, therefore, prefer a
slab corona geometry in terms of polarization.

\begin{figure*}[h]
    \centering
    \subfigure{\includegraphics[width=0.45\textwidth]{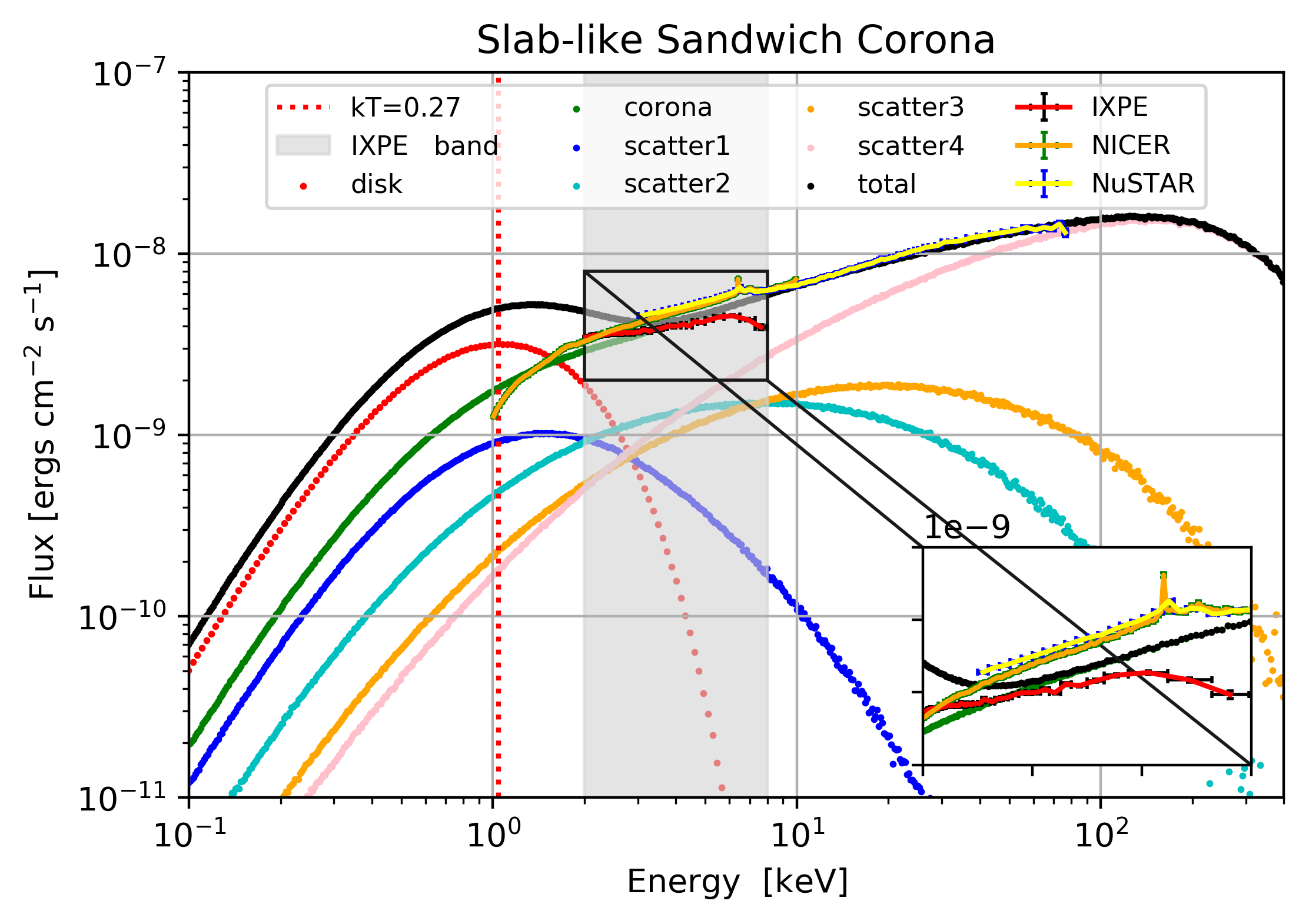}}
    \subfigure{\includegraphics[width=0.45\textwidth]{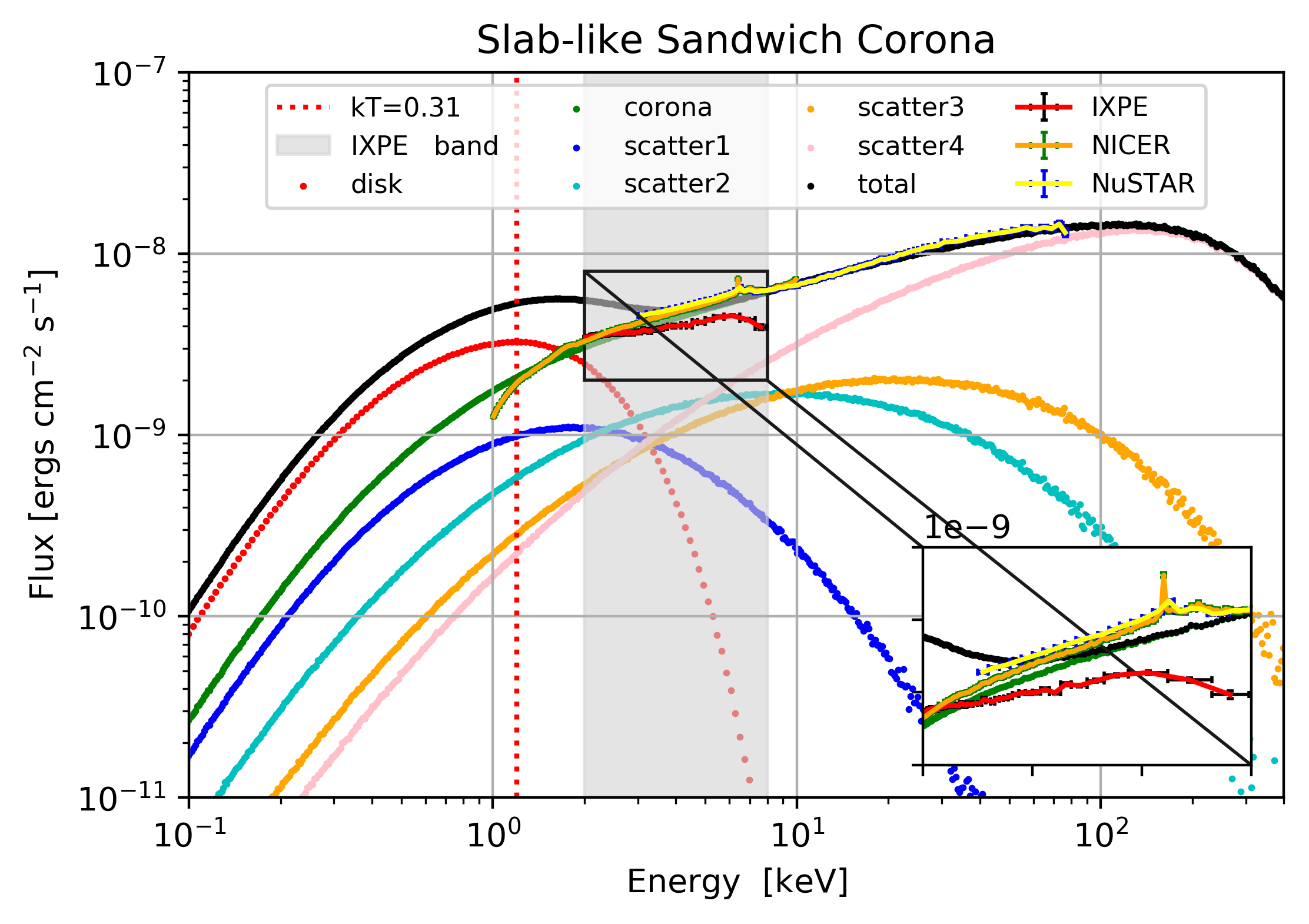}}
    \subfigure{\includegraphics[width=0.45\textwidth]{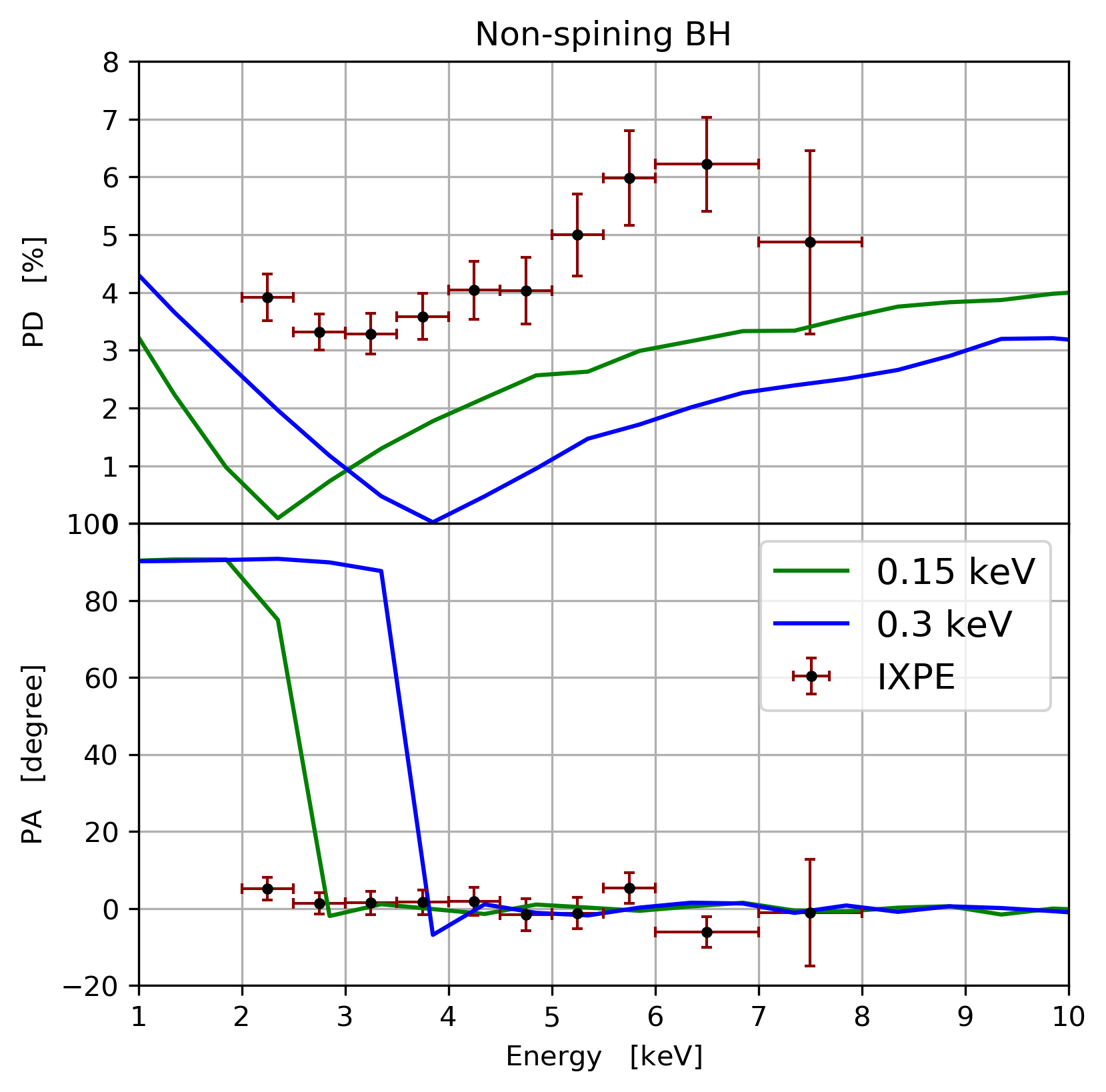}}
    \subfigure{\includegraphics[width=0.45\textwidth]{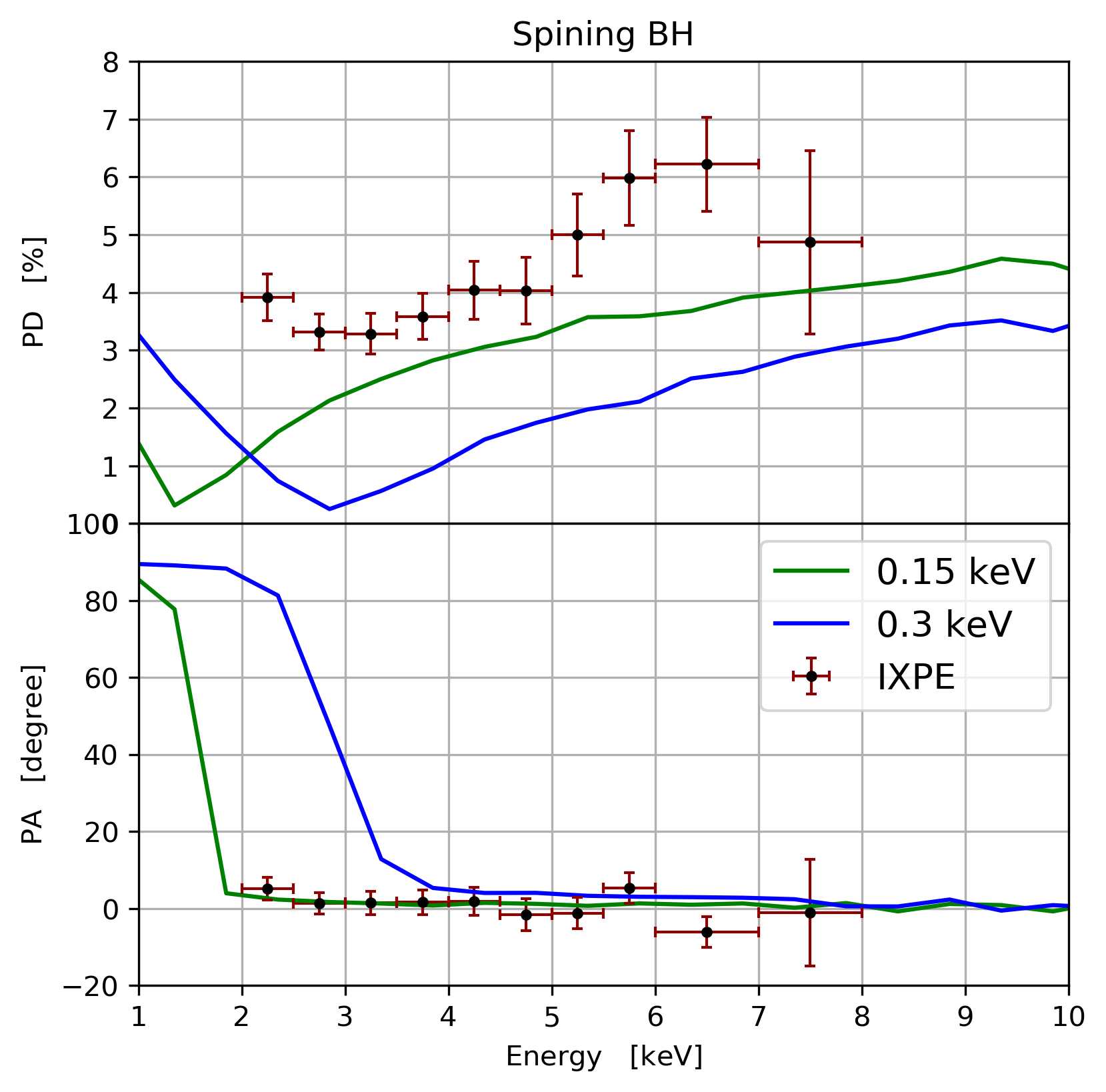}}
    \caption{\textbf{The simulated spectrum with disk temperature 0.3 keV and the comparison between
    the simulated and observed polarization properties.} \textcolor{black}{We adopt the parameters shown
    in the left column (sandwiching slab corona) of table~\ref{simp} but increased the accretion rate
    by a factor of 20 so that the simulated disk temperature is approximately 0.3 keV.} The upper
    panel shows the spectra of sandwiching slab corona both Schwarzschild (left) and Kerr (right) BH
    with $a = 0.9$ cases, respectively. We adopt the same color scheme as in figure~\ref{sim-spec} for
    consistency. The lower panel compares the polarization properties between the 0.15 keV (green) and
    0.3 keV (blue) cases for both Schwarzschild (left) and Kerr (right) BH, as well as the
    \textit{IXPE} observed polarization. 
    {Alt text: Two simulated spectrum and two comparison between polarization plots for hard state of
    Cyg X-1.}}
    \label{sim-compare}
\end{figure*}

\textcolor{black}{The main difference between the one- and two-Comptonization components model is that
the soft component decreases the inner disk temperature by a factor of two, from 0.3 to 0.15 keV.
Since the sandwiching slab corona model fits the observed polarization data better, we compared the
polarization properties between the disk temperature 0.15 keV case and 0.3 keV case again with
$\iota=60^{\circ}$, as summarized in figure~\ref{sim-compare}.}

The upper panel of figure~\ref{sim-compare} shows the spectrum with a disk temperature of 0.3 keV
for both the Schwarzschild and Kerr cases. In both cases, the disk emission is large and dominates
the low-energy part of the \textit{IXPE} band, resulting in larger discrepancies between the observed
and simulated spectra compared to the $kT=0.15$ keV cases (shown in the lower panel of
figure~\ref{sim-spec}). For $kT = 0.15$ keV cases, photons that were scattered at least four times
contribute predominantly to the total emission, almost throughout the \textit{IXPE} energy band. In
contrast, for $kT = 0.3$ keV, such multiply scattered photons dominate only at high-energy, leading
to distinct changes in the polarization signatures. They are illustrated in the lower panel of
figure~\ref{sim-compare}, which compares the polarization properties between the two disk
temperatures. We find that PD increases by \textcolor{black}{approximately 2 percentage points in the
\textit{IXPE} 2--8 keV band} when the disk temperature decreases from 0.3 keV to 0.15 keV, in the
Schwarzschild and Kerr geometries. Moreover, PD increases by about \textcolor{black}{1 percentage
point in the \textit{IXPE} energy band} when the BH spin increases from 0 to 0.9.
\textcolor{black}{However, the simulated PDs are still significantly lower than the observed ones
even with a rather high $60^{\circ}$ inclination.} In both temperature cases,
the PD exhibits a clear energy dependence. This is because photons scattered at least four times
increasingly dominate the high-energy spectrum, enhancing PD at higher energy. The depolarization
between photons that were scatted once or twice and those that were scattered three or more times is
due to their PAs differing by almost $90^{\circ}$ and dominating the low-energy bands. Therefore, the
net PD decreases in the low-energy band and finally increases in the high-energy band, presenting an
apparent energy dependence. The simulated PD with a disk temperature of 0.15 keV and a spin parameter
of 0.9 partially explains the detected high PD. Moreover, the simulated PA with the same model
configuration fits the observed PA well. However, the discrepancy between the observed high PD and
the theoretical prediction is still not fully solved. And we need further investigations in more
detail, which we will do in future works.

\textsc{MONK} does not currently allow us to tune the disk emission independently from the corona.
Although the contribution of disk emission is hence larger than observed for the 0.3 keV case, the
overall trends in polarization we discussed remain robust.

Then, let us also discuss how the polarization properties depend on other parameters. The BH
spin parameter $a$ and the optical depth $\tau$ affect the photon index of the spectrum
(larger $a$ and $\tau$ resulting in a harder spectrum). The PD remains almost the same between
the non-spinning BH and the spinning BH for the wedge-shaped corona, but $30\%$ larger in the
\textit{IXPE} band for the spinning BH of the slab corona case, but still needs inclination
higher than binary. The disk is close to the BH and rotates rapidly for the slab corona. Then, the
photons will suffer from stronger special and general relativistic effects, resulting in a larger PD. 

So far, we have omitted the reflection components in our simulation.
The reflection components are highly polarized (with PD reaching $\sim25\%$) and can
be strong in a lamp-post geometry (\cite{Dovciak2011}). In contrast, in a slab corona geometry, the
reflection components are suppressed by the hot corona, resulting in a relatively low reflection
intensity. The reflected photons behave similarly to the multiply scattered photons and enhance the
total intensity due to the effective increase of the number of scattering events in the corona, with
PA typically oriented perpendicular to the disk. When accounting for the relative flux, the contribution
of the reflection components to the net PD is around $2\%$ in the lamp-post geometry and even lower
($\sim1\%$) in the slab corona geometry (\cite{Schnittman2010}). Since disk reflection does not
remarkably affect the polarization in the \textit{IXPE} energy band, for simplicity, we neglect
the reflection components in our simulation. We will investigate the impact of reflection components
in detail in future studies.

We also ran the simulation using another tool RAIKOU (\cite{Kawashima2023}),  where the
polarization effects have been implemented in (Komine et al. in prep.), with a polarization
extension and confirmed these properties, for particular the sandwiching slab
corona. Specifically, we reproduced the energy dependence of PD (the dip below 1 keV and the
positive energy dependence in the IXPE band) and constant PA. RAIKOU yields a larger PD, nearly
twice as that obtained by MONK, which we believe is due to the different seed photon spectrum;
with RAIKOU we simulated the diskBB-like spectrum where the temperature is non-uniform and
the hottest regions are spatially localized, resulting in a higher PD. The conclusion that a
lower disk
temperature produces a larger PD remains unchanged. Since the observed energy dependence of PD
agreed with the simulation both by MONK and RAIKOU, we can safely conclude that the Cyg X-1 data
observed in 2022 May prefer a sandwiching slab corona geometry rather than a wedge-shaped corona
with truncated disk. We will investigate the polarization properties of two
Comptonization components in detail with more accurate simulations (e.g., diskBB seed photon
distribution, and reflection components taken into account).

\section{Conclusion}

We analyzed the May observations of Cyg X-1 in the low hard state seen by \textit{IXPE},
\textit{NuSTAR}, and \textit{NICER} in 2022. We observed significant variations in the light curves
of observation 1 and selected data of 8 time periods when \textit{IXPE} and \textit{NICER} observed
Cyg X-1 simultaneously. We fit those data by the one-Comptonization component model and the
two-Comptonization components model and found that the latter can better reproduce the data with
reasonable parameters. The variation is mainly attributed to the changes of the soft component. 

We divided observation 1 data based on the hardness ratio into the relatively hard and
soft parts and applied model independent polarimetric analysis. We found an obvious energy
dependence of PD in both parts. Simultaneous fitting of \textit{NICER} and \textit{IXPE}
suggests an inner disk temperature of about 0.15 keV, which is about half of that presented
in the discovery paper. The low disk temperature leads to a high observed PD, since the
photons need to encounter larger numbers of scatterings to reach the \textit{IXPE} band. The
calculated optical depths of the relatively hard and soft Comptonization components are similar
to those observed in 2005, indicating that Cyg X-1 in 2022 was in a typical hard state.

We investigated the spectrum and polarization properties based on the MONK code simulation
for a wedge-shaped corona with a truncated disk and a slab corona sandwiching the disk,
respectively. The wedge-shaped corona exhibits an almost constant PD in \textit{IXPE} band and
shows the energy dependence of PA. The slab corona presents an apparent energy dependence
of PD with an almost constant PA. PD will increase by \textcolor{black}{about 1--2 percentage points}
when the disk temperature decreases from 0.3 keV to 0.15 keV, due to the increase in the number of
scatterings to reach the \textit{IXPE} band. It will also increase if the BH is spinning
for a slab corona. \textcolor{black}{The simulated PDs are still significantly lower than the
observed ones with a rather high $60^{\circ}$ inclination.} Lowering the disk temperature from
0.3 to 0.15 keV reduces the discrepancy between the observed and theoretical PD, but it is not
fully solved.

In short, we conclude that the data of Cyg X-1 observed in 2022 May prefer a two-Comptonization
corona of different optical depth with a disk temperature of about 0.15 keV, and the effects of
the two-Comptonization components model will be studied in the future simulation work.

Lastly, the two-Comptonization components model may also be applied to other sources
such as IGR J17091-3624 (IGR J17091). \citet{Debnath2025} analyzed the spectrum and polarization
properties with data simultaneously observed by \textit{IXPE} and \textit{NuSTAR} in March, 2025.
They reported a high PD ($\sim8\%$) on average and an obvious energy dependence in PD, with an
inclination of $\sim60^{\circ}$. They attributed this high PD qualitatively to multiple scatterings
in the corona and strong relativistic effects of high-energy seed photons ($kT_{in} \sim0.4$ keV). We
quantitatively evaluated such an effect and obtained a high PD ($\sim 5\%$) as shown in
figure~\ref{sim-compare}. We should be reminded that our predicted PD does not fully explain the
observed high PD of the source. The low electron temperature obtained by their spectral fitting may
reduced the discrepancy reminded.

\section{Funding}
    This work was supported by the China Scholarship Council (CSC) under Grant No. 202006040056.
    Part of this work was supported by the JSPS KAKENHI Grant Numbers 23K25882, 23H04895 (T.M.) 
    21H04488 (K.O.), JP22H00128,  JP23K03448, JP24K00672 (T.K.) and JP23H00117 (H.T. and T.K.).
    Test computations for the discussion with RAIKOU code were carried out on the XD2000 at the Center
    for Computational Astrophysics, National Astronomical Observatory of Japan.


\end{document}